\documentclass[manuscript,screen,nonacm]{acmart}
\AtBeginDocument{%
  }
\setcopyright{acmcopyright}
\copyrightyear{2018}
\acmYear{2018}
\acmDOI{XXXXXXX.XXXXXXX}

\acmJournal{IMWUT}
\acmVolume{37}
\acmNumber{4}
\acmArticle{111}
\acmMonth{8}

\usepackage{graphicx}
\usepackage{subfig}
\usepackage[ruled]{algorithm2e}
\usepackage{caption}
\usepackage{color}

\newcommand{\cg}[1]{{\color{black}{#1}}}%
\newcommand{\rev}[1]{{\color{black}{#1}}} %

\begin{document}

\title{DisPad: Flexible On-Body Displacement of Fabric Sensors for Robust Joint-Motion Tracking}
\author{Xiaowei Chen}%
\email{wdenxwa@stu.xmu.edu.cn}

\affiliation{%
  \institution{Xiamen University}
  \streetaddress{No. 422, Siming South Road}
  \city{Xiamen}  \state{Fujian}
  \country{China}
  \postcode{361005}
}

\author{Xiao Jiang}
\email{24320182203214@stu.xmu.edu.cn}
\affiliation{%
  \institution{Xiamen University}
  \streetaddress{No. 422, Siming South Road}
  \city{Xiamen}  \state{Fujian}
  \country{China}
  \postcode{361005}
}

\author{Jiawei Fang}
\email{22920202204553@stu.xmu.edu.cn}
\affiliation{%
  \institution{Xiamen University}
  \streetaddress{No. 422, Siming South Road}
  \city{Xiamen}  \state{Fujian}
  \country{China}
  \postcode{361005}
}

\author{Shihui Guo}

\email{guoshihui@xmu.edu.cn}
\authornote{Corresponding author: guoshihui@xmu.edu.cn.}
\affiliation{%
  \institution{Xiamen University}
  \streetaddress{No. 422, Siming South Road}
  \city{Xiamen}  \state{Fujian}
  \country{China}
  \postcode{361005}
}

\author{Juncong Lin}%
\email{jclin@xmu.edu.cn}

\affiliation{%
  \institution{Xiamen University}
  \streetaddress{No. 422, Siming South Road}
  \city{Xiamen}  \state{Fujian}
  \country{China}
  \postcode{361005}
}

\author{Minghong Liao}%
\email{liao@xmu.edu.cn}

\affiliation{%
  \institution{Xiamen University}
  \streetaddress{No. 422, Siming South Road}
  \city{Xiamen}  \state{Fujian}
  \country{China}
  \postcode{361005}
}

\author{Guoliang Luo}%
\email{luoguoliang@ecjtu.edu.cn}

\affiliation{%
  \institution{East China Jiao Tong University}
  \streetaddress{No.150 Shuanggang Road}
  \city{Nanchang}
  \state{Jiangxi}
  \country{China}
  \postcode{330013}
}

\author{Hongbo Fu}%
\email{hongbofu@cityu.edu.hk}

\affiliation{%
  \institution{City University of Hong Kong}
  \streetaddress{Tat Chee Avenue}
  \city{Hongkong}
    \state{Hongkong}
  \country{China}
  \postcode{999077}
}

\renewcommand{\shortauthors}{Xiaowei et al.}

\begin{abstract}
The last few decades have witnessed an emerging trend of wearable soft sensors; however, there are important signal-processing challenges for soft sensors that still limit their practical deployment. They are error-prone when displaced, resulting in significant deviations from their ideal sensor output. In this work, we propose a novel prototype that integrates an elbow pad with a sparse network of soft sensors. Our prototype is fully bio-compatible, stretchable, and wearable. We develop a learning-based method to predict the elbow orientation angle and achieve an average tracking error of 9.82 degrees for single-user multi-motion experiments. With transfer learning, our method achieves the average tracking errors of 10.98 degrees and 11.81 degrees across different motion types and users, respectively. Our core contributions lie in a solution that realizes robust and stable human joint motion tracking across different device displacements.
\end{abstract}

\ccsdesc[500]{Human-centered computing~Ubiquitous and mobile computing systems and tools}
\ccsdesc[500]{Human-centered computing~Ambient intelligence}

\keywords{soft sensors, textile sensors, motion tracking, robust signal processing, long short-term memory, transfer learning, domain adaption, fuzzy entropy}
\maketitle

\section{Introduction}
\begin{figure}[t]
    \centering
    \includegraphics[width=0.9\columnwidth]{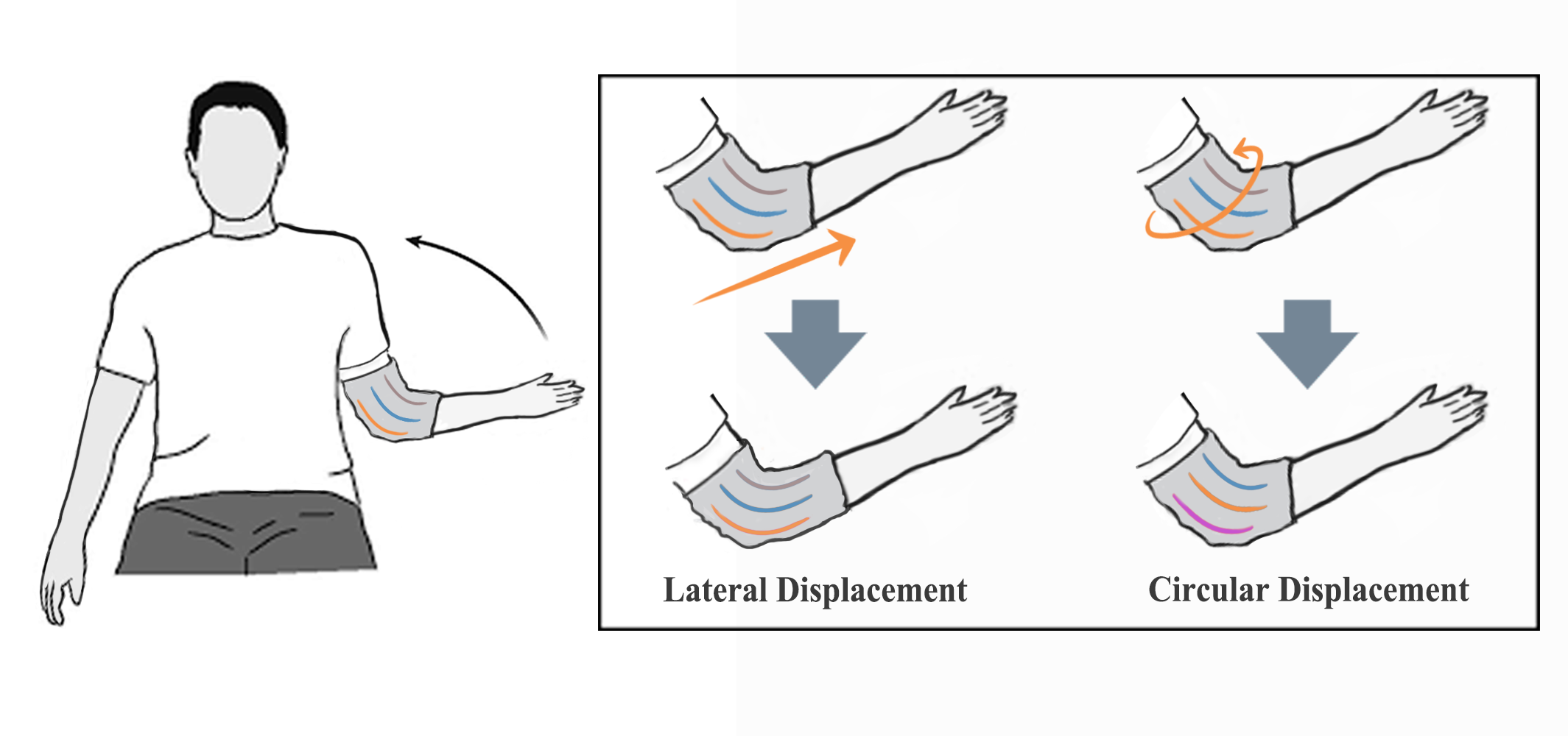}
      \caption{Our prototype, DisPad, can robustly track human joint orientation despite lateral and/or circular sensor displacement. This work uses the elbow to exemplify the function of robust joint tracking of our method.}~\label{fig:HumanDisplacement}
\end{figure}

Human motion tracking has been actively explored not only regarding its great potential for understanding human intentions \cite{kale2016study}, but also for actively controlling physical devices and systems \cite{chen2012sensor,schneegass2015simpleskin,schneegass2014towards}. 
As a new type of human-computer interaction, human motion tracking could be applied in diverse fields, including robotics, haptics, bio-mechanical studies, rehabilitation, health care, and entertainment \cite{chossat2015wearable,cornacchia2016survey}.
The human motion tracking solutions widely used in the film, animation, and gaming industries are based on optical tracking \cite{huang2018deep} or inertial sensors \cite{kunze2014sensor}. 
Some studies have also proposed solutions to combine these two types of sensors \cite{ahuja2021pose}. 
These solutions have achieved great success for professional applications, but may not be ideal for consumer-level applications, prioritizing comfort and convenience of user experience. 
Soft sensors are emerging as promising solutions for monitoring human activities \cite{voit2017fabricid,poupyrev2016project}. Their advantages in terms of bio-compatibility, high stretchability, lightweight, and ease of integration within clothing allow long-term monitoring of human physical status.
With soft sensors, motion capture is not easily affected by the aforementioned factors, such as lighting conditions, integration drift, and occlusions. 
Thus, there has been a growing demand for applying soft wearable sensors to the domains of human motion capture, human-computer interfaces, and soft robotics, amongst others \cite{amjadi2016stretchable}. 

Soft 
sensors face critical challenges before fulfilling their potential for wide HCI applications.
One of the challenges is device placement \rev{(we refer to them as DisPad position below)}, which \rev{also varies} significantly across different users, motion types, and wearing sessions (Fig.~\ref{fig:HumanDisplacement}).
However, existing methods require expert knowledge to strictly maintain
the device location on a subject~\cite{kunze2008dealing,kunze2014sensor,someya2019toward}, hampering system robustness and convenience.
In specific scenarios (such as stroke patients with motor impairment), such a requirement of fixed position cannot be rigorously applied. 
{Thus, the prediction models based on the assumption of a fixed position of device placement would not be practical in real-use scenarios and may produce serious measurement errors when such an assumption does not hold.}

To address these challenges, we present \emph{DisPad} for tracking human motion accurately and robustly, even in the context of considerable variations in device placement. As a proof of concept, we select the elbow joint given the popularity of athletes using pads to protect them against injury during a fall or strike. We believe with appropriate modification to our device and procedure, our learning-based approach can also be effectively applied to other joints. 
DisPad builds upon a standard textile elbow pad and a sparse network of six stretchable sensors.
Our approach removes the strict constraints of accurate device placement on the body, allowing users to freely wear our prototype in various configurations, just like a regular pad.
The flexible on-body configuration considerably affects the raw sensor readings and presents significant challenges for sensor signal processing and task-relevant algorithms.
Our core contributions lie in a solution that realizes low-cost, robust, and stable human joint motion tracking across different device displacements. 
To achieve this, we present a learning-based algorithm to predict the joint angle {in a sagittal plane} based on sensor readings and introduce a technique of transfer learning to further address the variations caused by different users and motion types.
Our method offers users the flexibility of wearing the pad with minimal constraints and ensures our system's maximal convenience. 
In sum, the contributions of our work are three-fold. 
\begin{itemize}
    \item A learning-based method to handle the variation in sensor signals caused by sensor displacement and to achieve the goal of flexible on-body device placement. 
      We use a long short-term memory (LSTM) to predict the joint angle, achieving an average error of 9.82 degrees on the testing dataset, where the elbow pad is placed randomly within a large region on the same user's arm.  
      \item An \cg{unsupervised method based on fuzzy entropy and transfer learning to adapt the model to datasets with different feature distributions. It %
      achieves \rev{an} average error %
      of 10.98 degrees across different motion types and 11.81 degrees across different users.} 
    \item A system prototype including sensor signal collection, transmission, angle prediction, and graphical visualization to predict human arm joint angle.
    As a complete solution comprised of both software and hardware, this system proves the potential of using soft sensors that can be flexibly worn for robust motion-tracking purposes.
\end{itemize}

\section{Related Work}

\subsection{Human Motion Capture}

Existing methods for human motion capture include vision-based \cite{gioberto2012garment,chen2007real,kale2016study} and non-vision-based approaches \cite{lee2016novel,kiaghadi2018fabric,gibbs2005wearable}. 
Vision-based methods use RGB, RGB-D, or infrared cameras to capture images and identify human posture.
Benefiting from recent success of deep learning, vision-based methods have achieved accurate body tracking under ideal conditions (sufficient lighting and no occlusion).
Typical non-vision-based methods attach sensors, such as IMU \cite{steven2018feature} or ultrasonic sensors \cite{einsmann2005modeling}, to a human body to measure the orientation and position of specific body parts \cite{zhang2015whole}.
As IMU sensors are low-cost and small size, IMU-based tracking systems are so far the state-of-the-art wearable solutions for unconstrained indoor and outdoor environments \cite{steven2018feature,tian2015upper}. 
To achieve accurate and stable tracking results, researchers developed algorithms to handle the drift problem caused by the integration operation  \cite{xiao2018machine,zhang2015whole,von2018recovering,filippeschi2017survey}. 
For example, Yizhai et al. used an Extended Kalman Filter (EKF)-based algorithm to rectify drifts \cite{zhang2015whole}.
Existing vision-based and non-vision-based solutions have been widely applied in a variety of professional and consumer applications. 
Wearable systems with flexible sensors complement existing methods and extend the boundary to scenarios where users demand wearing comfort over long sessions and convenience to use in both indoor and outdoor scenarios.

Researchers have already explored the use of soft sensors for motion tracking.
To date, existing methods have explored the applications of tracking the motion of the upper body \cite{farringdon1999wearable}, fingers \cite{lorussi2005strain,bae2013graphene,chossat2015wearable,glauser2019interactive}, lower limbs \cite{gibbs2005wearable,mengucc2014wearable}, elbow joints \cite{meyer2006textile}, and knee joints \cite{tognetti2014new}.
In addition to tracking joint angles, a dense array of soft sensors can be used to measure localized area changes and reconstruct complex skin deformations \cite{glauser2019deformation} and tiny stretches of skin (by pulsing of arteries) \cite{gong2015highly}. 
Some researchers have combined soft sensors with inertial sensors to achieve joint motion tracking  \cite{yirmibesoglu2016hybrid}. 
Yirmibesoglu et al. developed a hybrid soft sensor for measuring motion \cite{yirmibesoglu2016hybrid}, made of a hyper-elastic silicone elastomer containing embedded micro-channels filled with conductive liquid metal. 

The existing studies in this field have mainly focused on the metrics of accuracy and real-time performance of the sensors for human motion capture. 
A recent study applied fabric sensors to track joint angles while considering a small range of sensor displacement ($\leq$1 cm) \cite{liu2019reconstructing}. 
Since this prior work confirms the effect of sensor displacement (even within a moderate range) on tracking accuracy, our work presents a thorough investigation of this open question and explores potential solutions to achieve accurate tracking under conditions of large sensor displacement.

\subsection{Robust Sensor Placement}

The problem of sensor displacement indicates that sensors may deviate from their ideal configuration and end up positioned in a significantly different configuration.
This issue often occurs when using wearable systems to track human states and poses challenges to pre-trained pattern recognition models.
This issue is exacerbated when using flexible sensors, as in our case, due to their deformable property.

One solution is to analyze displacement data and use algorithms to rectify the results.
One group of researchers assumed that the changes in sensor position and sensor feature distribution are consistent with covariant shift \cite{sugiyama2007covariate}. 
Based on this assumption, they proposed an improved cross-validation technique called material-weighted cross-validation, which can be used for model and parameter selection in classification tasks. 
Another study combined information from a gyroscope and an accelerometer to distinguish rotation and translation \cite{kunze2008dealing}. The authors proposed a heuristic algorithm to calculate the transformation matrix and classify human motion.
Foster et al. proposed an online self-calibration method to adjust a classifier model dynamically \cite{forster2009unsupervised}. 
Their method shows robust performance under circular displacement within 90 degrees, and our method extends to both lateral and circular displacement (within 360 degrees).
Another study used functional principal component analysis to compensate for the data changes caused by the positioning changes on the sensor body \cite{haratian2013toward}. %
One study maximized expectations and estimated feature distribution in an unsupervised manner, creating a mechanism for estimating distribution offsets and adjusting the original classifier \cite{chavarriaga2013unsupervised}. 
The aforementioned works \cite{forster2009unsupervised,haratian2013toward,chavarriaga2013unsupervised} consistently reveal that the performance degraded as the degree of displacement became larger. 
Banos et al. used a hierarchical weighted classifier to dynamically search for available features \cite{banos2014dealing}. 
\begin{figure}[b]
\subfloat[]{\includegraphics[width=2.4in]{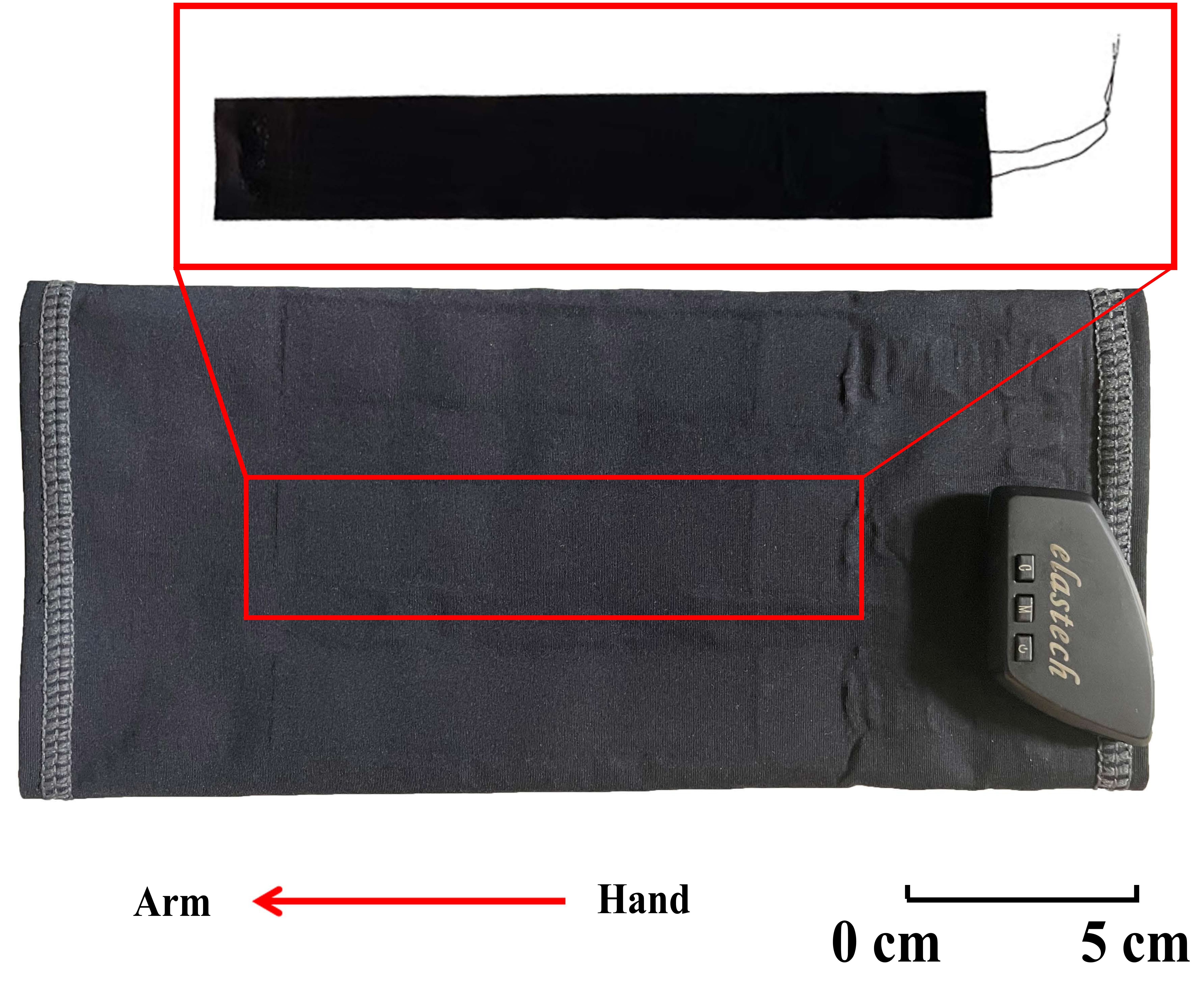}%
\label{fig:elbowpad}}
\hfil
\subfloat[]{\raisebox{3ex}%
{\includegraphics[width=1.5in]{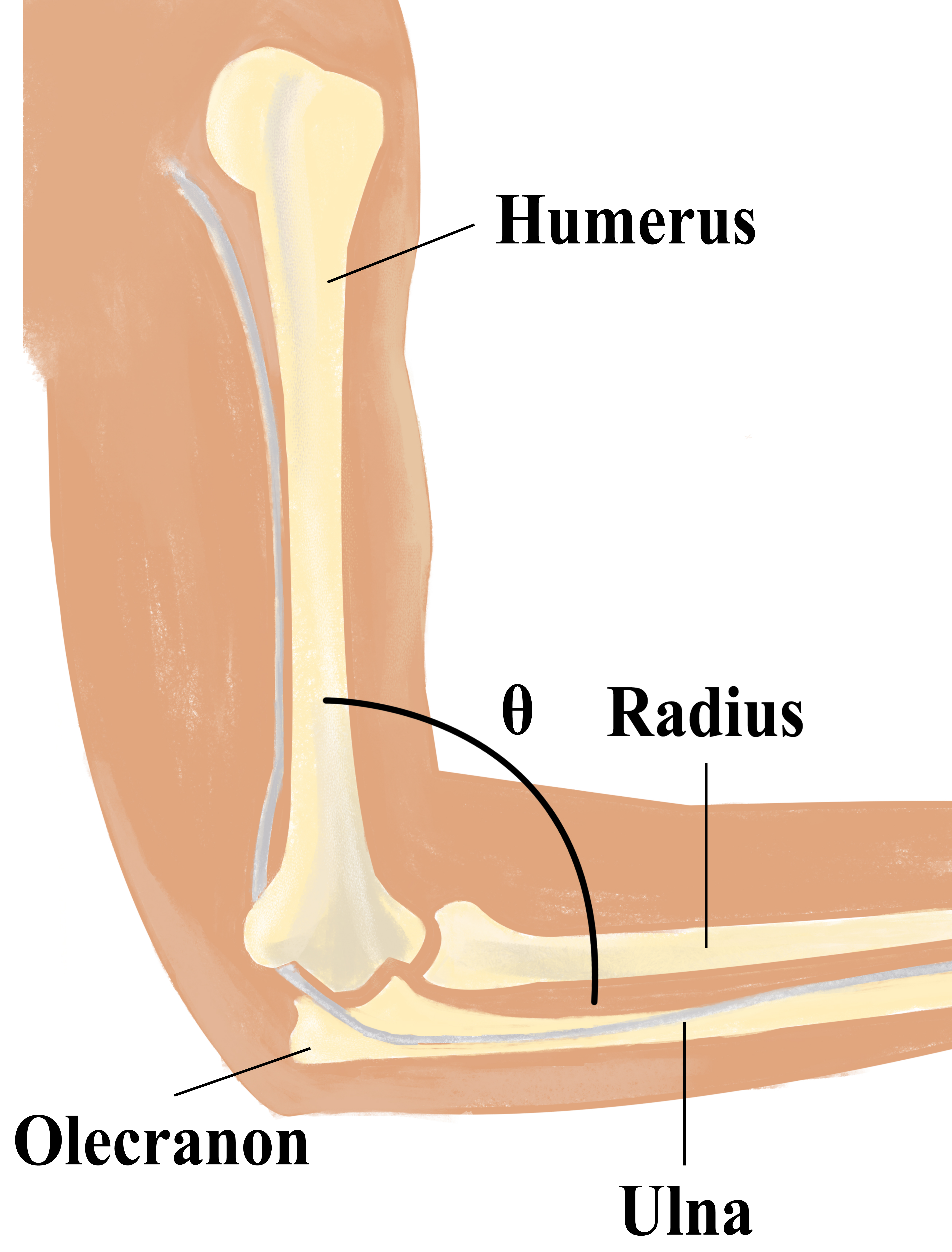}}%
\label{fig:arm}}
\caption{(a) The appearance of DisPad, with a red arrow implying the wearing direction. (b) The illustration of the bending angle of an elbow joint.}
\label{fig:ElbowArm}
\end{figure}
Some researchers tend to search for features that are independent of sensor position. Forster et al. adopted a genetic algorithm to extract features independent of location shifts for the task of motion type classification \cite{forster2009evolving}. 
Another solution is to increase the sensing area \cite{enokibori2014human}. 
The sensing is more robust in the situation of sensor displacement because of the enlarged sensing area. 
To sum up, sensor displacement has been identified as one of the critical factors causing %
performance degradation, and existing works have explored a variety of solutions to resolve this issue.
Due to the deformation characteristics of soft sensors, it is more complicated to achieve robust pattern recognition in the case of placement deviation.
Our method aims to tackle the challenges of large circular and lateral device displacements of an elbow pad and gain robust joint tracking performance.

\section{Methodology}
\rev{Let $\mathcal{D}_{SS}$ be the dataset containing data from a single user who collected sensor data and corresponding ground truth at various DisPad positions, $\mathcal{D}_{SM}$ be the dataset including data from the same single user above who performs four different motions at only three DisPad positions, $\mathcal{D}_{MM}$ be the dataset containing data from ten users who perform random motions at only three DisPad positions, $\mathcal{D}_{NM}$ be the dataset including data from one motion of $\mathcal{D}_{SM}$ but without labels, $\mathcal{D}_{NU}$ be the dataset including data from one user of $\mathcal{D}_{MM}$ but without labels, we aim to infer human joint bending angle $\theta$ while there is sensor displacement and generalize this method to other users and motions. Here we define $\mathcal{D}_{SS}$ as the source domain, $\mathcal{D}_{NM}$ and $\mathcal{D}_{NU}$ as the target domain.}

\subsection{Overview}

Our work designs and develops a prototype, DisPad, based on a standard flexible pad and a sparse network of six soft and stretchable sensors (Fig.~\ref{fig:elbowpad}).
Our method aims to estimate the bending angle, $\theta$, of an elbow joint (Fig.~\ref{fig:arm}) from the sensor readings of the flexible sensors. \rev{Besides, for new users/motions, our method can robustly predict their moving angles without collecting the optical ground truth.}
We define the bending angle as the angle in the sagittal plane between the humerus and the central line between the radius and the ulna (Fig.~\ref{fig:arm}).

\cg{We will investigate this problem in two
different use scenarios. One is for a single user who collected \rev{sensor} data and \rev{corresponding ground truth} with various DisPad positions, and the other is for \rev{new} motions/users with a few random DisPad positions (Fig. \ref{fig:overview}). \rev{Note that the ground truth of new motions/users is unknown.}
For a single user, we first collect the sensor readings under various lateral and circular displacements. At the same time, an optical motion-capture (MoCap) system records the elbow bending angles as ground truth. Then, we train an
LSTM model by minimizing the \rev{MSE loss} between the ground truth and the estimated joint angle.%

\begin{figure*}[b]
    \centering
    \includegraphics[width=0.9\textwidth]{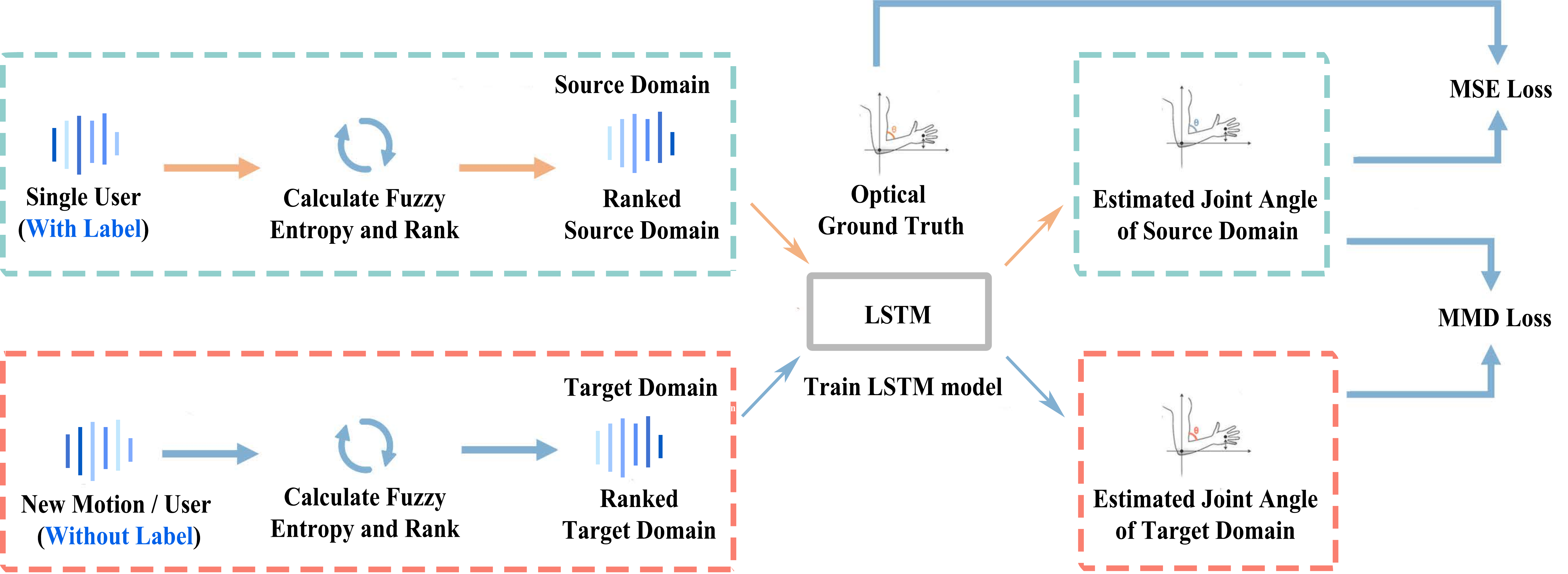}
      \caption{\cg{Method overview. %
      Note that we leverage six vertical lines with the varying color shading to visualize the magnitude of six sensor readings' information entropy.}} 
    \label{fig:overview}
\end{figure*}

To address the variations across different users and motion types, we design %
a calibration stage for new users and new motions. Specifically, we adopt %
a ranking method based on fuzzy entropy %
and transfer learning based on the LSTM model during this stage.
First, the fuzzy entropy value of every sensor reading during each displacement is computed to rank the different sensor readings according to the \rev{matrix} of the fuzzy entropy at every DisPad position. 
Second, the LSTM model undergoes the transfer learning process to mitigate the gap between new users or %
motion data and the original training dataset.
 Namely, the maximum mean discrepancy (MMD) loss is calculated from the output level (estimated joint angle) of the source domain and the target domain. %
Besides, the mean squared error (MSE) loss is calculated in the source domain to ensure the prediction ability of the neural network.}

\subsection{Unsupervised Transfer Learning for Calibration}

\cg{After we built the model based on $\mathcal{D}_{SS}$, we introduced the technique of unsupervised transfer learning to adapt our trained model to challenging conditions: unknown device displacement in multi-motion and multi-user scenarios, without the need to capture the ground truth using a professional MoCap system. 
This technique allows unsupervised calibration and guarantees the flexibility of the use of DisPad in real-world applications.

Given that $\mathcal{D}_{SM}$ and $\mathcal{D}_{MM}$ only contain a few DisPad positions, we decided to perform transfer learning %
on $\mathcal{D}_{SS}$ to $\mathcal{D}_{SM}$ and $\mathcal{D}_{MM}$. However, a direct transfer is hard as a user may wear DisPad with any displacement, making the possible DisPad positions on new users and new motions unknown. Thus, the transfer learning should handle both sensor displacement and different patterns of new motions/users. To address this problem, we employed sensor ranking to make the sensor signal perform consistently in every DisPad position first and used transfer learning to enable the model to predict $\mathcal{D}_{SM}$ and $\mathcal{D}_{MM}$ with different distributions. 
Namely, the calibration process is composed of two steps: 1) rank the sensor readings using the metric of fuzzy entropy; 2) refine the pre-trained model to an unknown condition of a DisPad position (transfer learning).} 
\begin{algorithm}[t]  
	\caption{DisPad: Unsupervised Transfer Learning}
	\label{alg::ag1}
	\LinesNumbered
	\KwIn{$\mathcal{D}_{SS}$, the data of a new user $\mathcal{D}_{NU}$ or a new motion $\mathcal{D}_{NM}$ at one DisPad position and $\mathcal{D}_{Train Source}=[]$}
	\KwOut{Moving angle of the new user or new motion $\theta$}
	\rev{$L^n=[l_1^n,l_2^n,...,l_6^n]\leftarrow$Indices of six sensor readings of $\mathcal{D}_{NU}/\mathcal{D}_{NM}$}\;
	Calculate the fuzzy entropy of $\mathcal{D}_{NM}/\mathcal{D}_{NU}$\;
	$\mathcal{D}_{Train Target} \leftarrow$Rank $\mathcal{D}_{NU}/\mathcal{D}_{NM}$ from small to large according to the fuzzy entropy matrix; the ranking index $\hat{L^n}=[\hat{l_1^n},\hat{l_2^n},...,\hat{l_6^n}$]\;
	\For{{every DisPad position of $\mathcal{D}_{SS}$}}{%
	$V^s=[V_1^s,V_2^s,...,V_6^s]\leftarrow$Values of six sensor readings of $\mathcal{D}_{SS}$ at this DisPad position\;
	$L^s=[l_1^s,l_2^s,...,l_6^s]\leftarrow$Indices of six sensor readings of $\mathcal{D}_{SS}$ at this DisPad position\;
	Calculate the fuzzy entropy of $V^s$\;
		$\hat{V^s}\leftarrow$Rank $V^s$ from small to large according to the fuzzy entropy matrix; the ranking index $\hat{L^s}$=[$\hat{l_1^s},\hat{l_2^s},...,\hat{l_6^s}$]\;
	\If{$[l_1^n,l_2^n]\cap[l_1^s,l_2^s]\neq \emptyset$}{
	$\mathcal{D}_{Train Target}\leftarrow\mathcal{D}_{Train Target}$.append($\hat{V^s}$)}
	}
	Input$\mathcal{D}_{Train Source}$ and $\mathcal{D}_{Train Target}$ to LSTM\;
	Train LSTM to compute MSE loss on $\mathcal{D}_{Train Source}$ and MMD loss between the estimated value of $\mathcal{D}_{Train Source}$ and $\mathcal{D}_{Train Target}$;
	
\end{algorithm}
{Alg.~\ref{alg::ag1} shows the pseudo-code of our method.}
\begin{figure}[t]
\subfloat[]{\raisebox{0ex}%
{\includegraphics[width=2.4in]{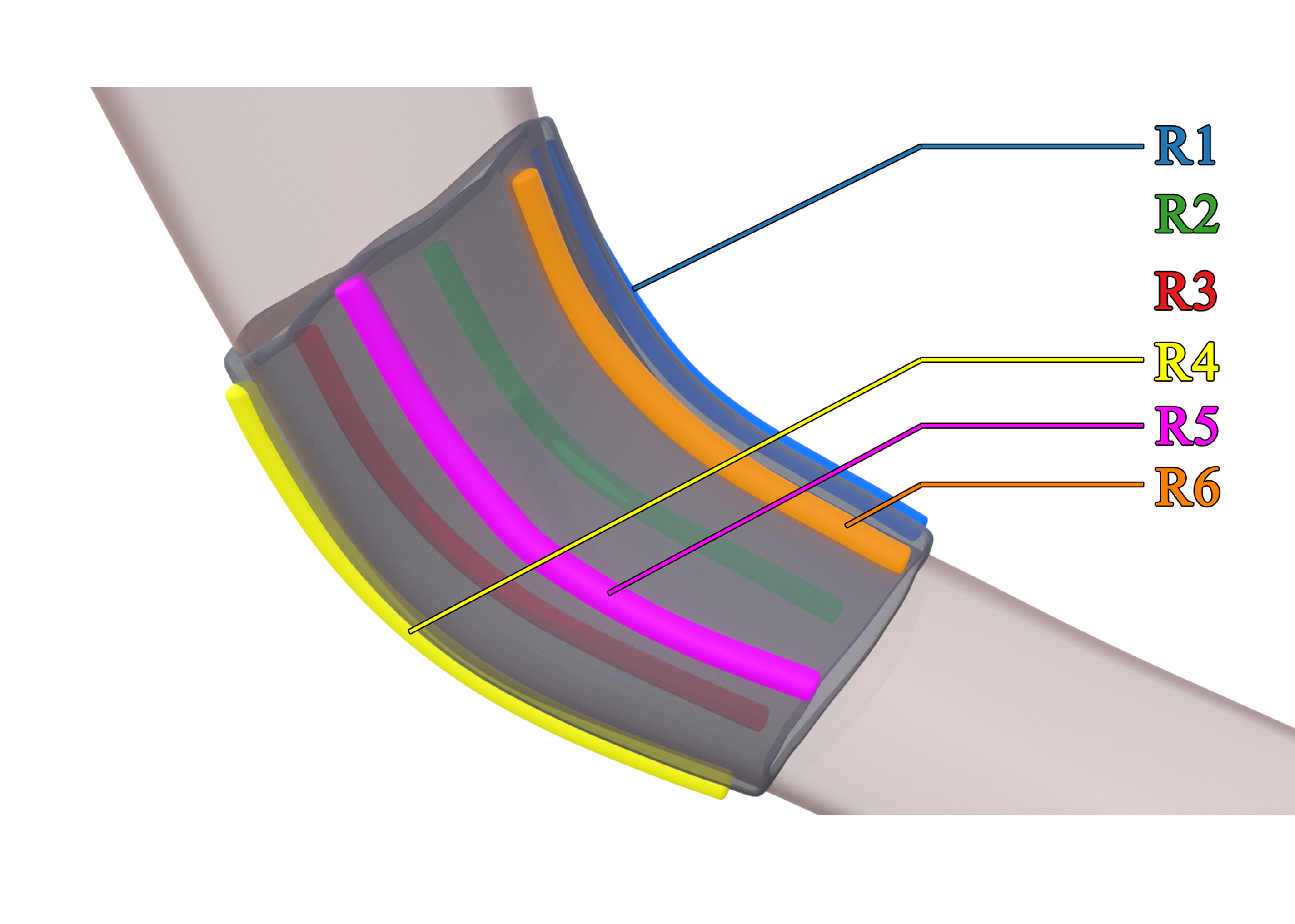}}%
\label{fig:iR2R3}}
\hfil
\subfloat[]{\includegraphics[width=2.4in]{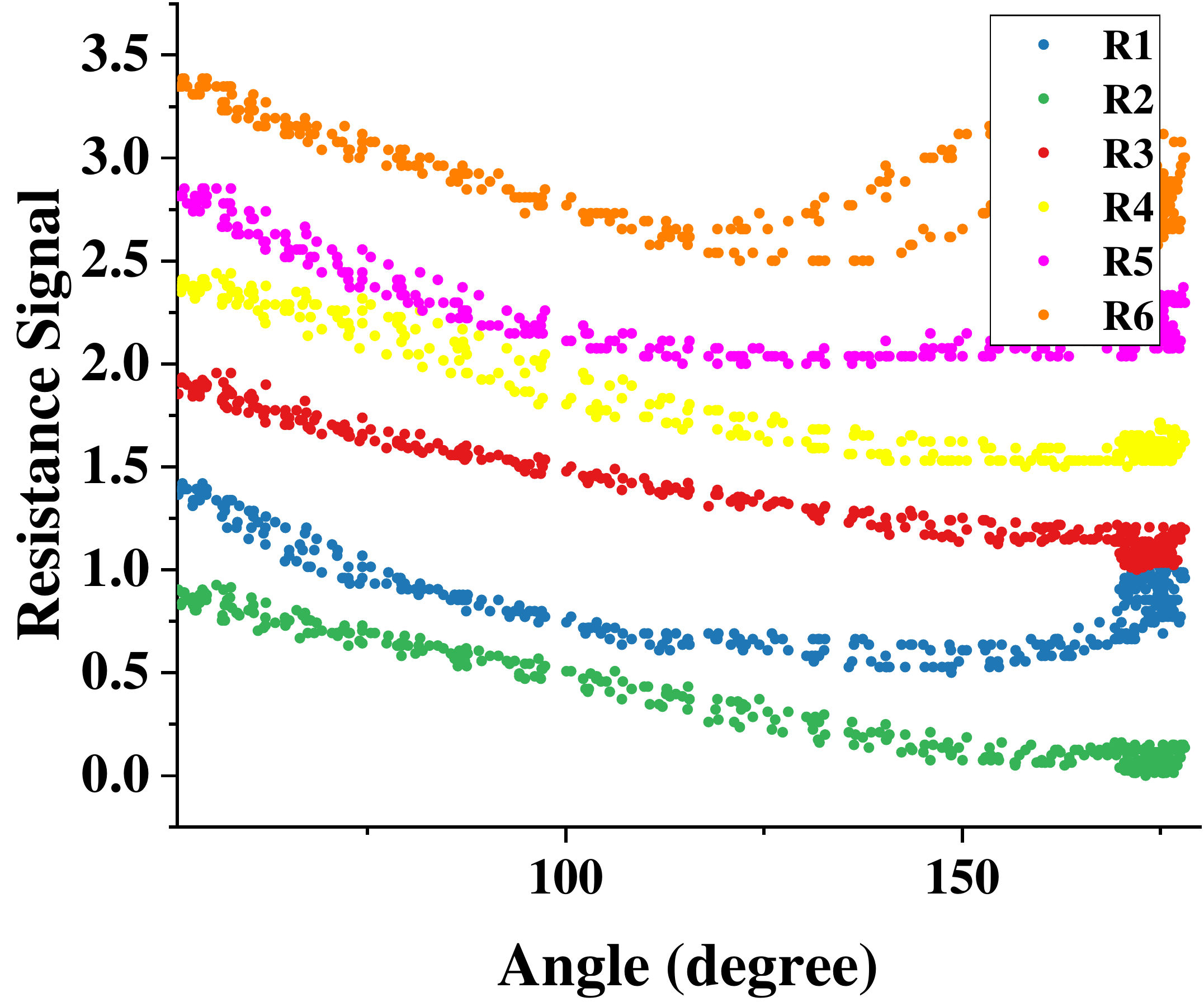}%
\label{fig:r2r3}}
\hfil
\subfloat[]{\includegraphics[width=2.4in]{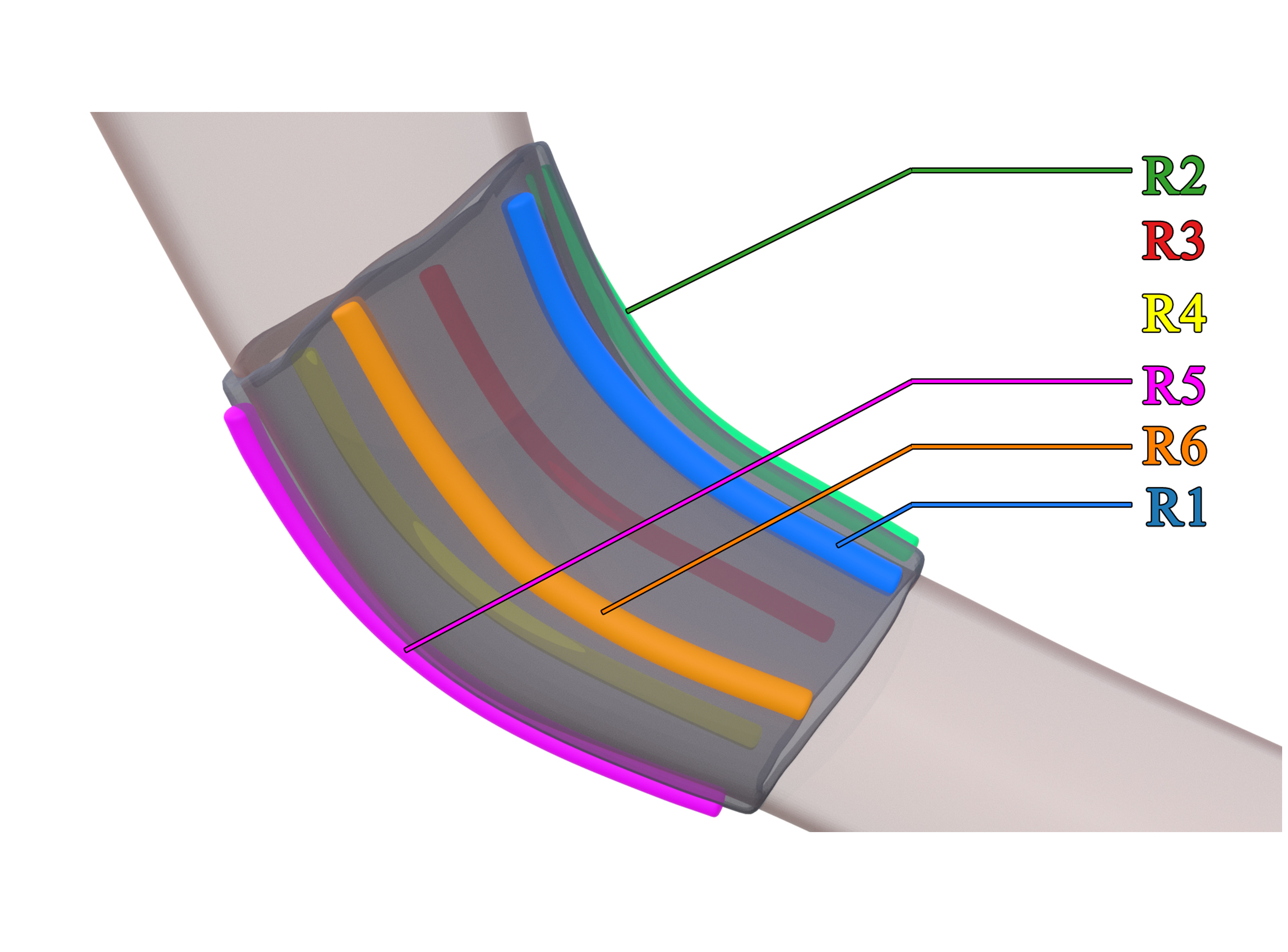}%
\label{fig:ir3r4}}
\hfil
\subfloat[]{\includegraphics[width=2.4in]{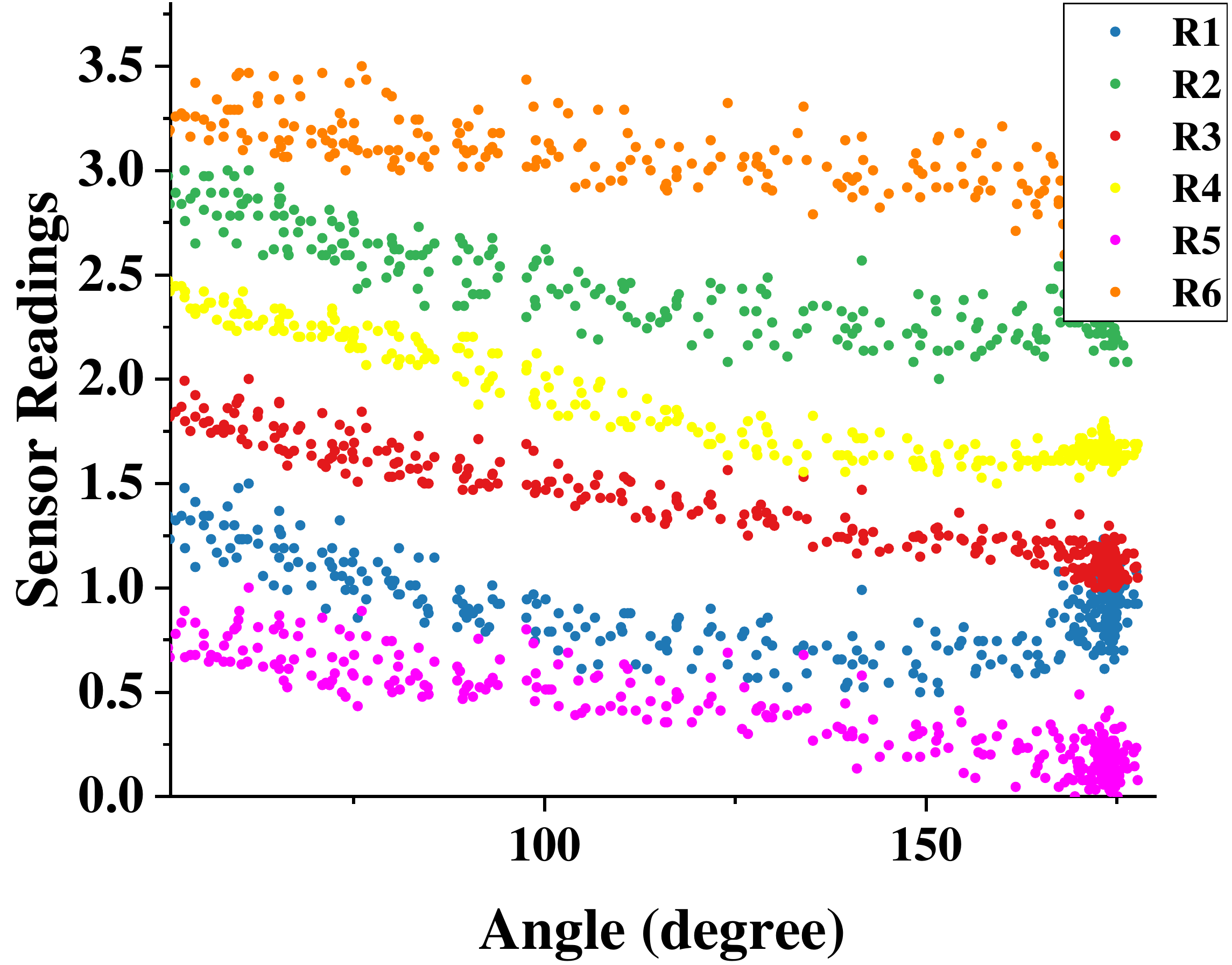}%
\label{fig:r3r4}}
\caption{The sensor readings change with different sensor placements. \rev{Note that we have normalized all the sensor readings and separated them by adding different scalars to make the figure clearer. }(a) A placement where R2 and R3 are on the side chelidon. Sensor readings correspond to the position (a). (c) A placement where R3 and R4 are on the side of the olecranon. 
    (d) Sensor readings correspond to the position (c).}
\label{fig:complexR}
\end{figure}
\subsubsection{Sensor Ranking with Fuzzy Entropy}
\rev{As Fig.~\ref{fig:r2r3} and~\ref{fig:r3r4} show,} the sensors not close to %
the side of 
the olecranon tends to show %
lower linearity as well as more noise compared with the ones close to the side of the 
olecranon.
Besides, It can be observed that the linearity of every sensor is different, and sensor readings with lower linearity are more likely to be accompanied by noise. Furthermore, the mappings between bending angles is more chaotic and produce one-to-more mappings. Thus, a sensor with lower linearity cannot strongly represent a user's motion pattern.
As the lateral and circular displacements cause sensors to change their positions, the sensor near the olecranon will also be different. As a result, the sensor representing the stronger motion pattern will change when there are sensor displacements. This change can confuse the network and make the prediction less effective. Thus, we propose to keep the sensor readings that are strongly representative of the motion pattern and the sensor data that are weakly representative of the motion pattern always in the same position by ranking operation. In that way, strongly representative sensor readings and weakly representative sensor readings are always in the same position. Therefore, the confusion of the network can be reduced. %
\cg{Since lower linearity and more noise represent the disorder of the sensor signal (that is, the possibility of {the signal repeating} {its previous pattern} is low), we decided to employ fuzzy entropy \cite{chen2007characterization} to measure the complexity of every sensor reading within at {every} DisPad position, {and then estimate whether the sensor can strongly represent a user's motion pattern.}

The fuzzy entropy algorithm proposed to analyze surface electromyography (EMG) signals \cite{chen2007characterization} measures the complexity of time series data. %
\cg{For a finite time series of \rev{one} sensor reading $\{x(i): 1 \leq i \leq N\}$ (size $N$), given the window size of $m$, the series can be \rev{formed} into vector sequences $\left\{\mathbf{x}_{i}^{m}, i=1, \ldots, N-m+1\right\}$, the fuzzy entropy can be estimated by the following equations:
\begin{align}
\label{E1}
\operatorname{Fuzzy} \operatorname{En}(m, r, N)&=\ln \phi^{m}(N, r)-\ln \phi^{m+1}(N, r), \\
\phi^m(N, r) &= \frac{1}{N-m+1} \sum_{i=1}^{N-m+1}  \frac{1}{N-m} \sum_{j=1, j\neq i}^{N-m+1} D_{ij}^m, \\
D_{ij}^m &= \exp \left(-\left(d_{\mathrm{ij}}^{m}\right)^{N} / r\right), 
\end{align}
where $r$ is the standard deviation of the original time series, and $d_{ij}^m$ is the maximum absolute difference between \rev{$\mathbf{x}_{i}^{m}$ and $\mathbf{x}_{j}^{m}$.}} 

{\rev{Through calculating the fuzzy entropy of each sensor, we could rank readings of different sensors according to the entropy matrix and use %
it as a new input. \rev{For six sensor readings at one DisPad position $X=[x_{1},x_{2},...,x{6}]$}, this procedure can be represented by:}}}
\begin{equation}
\label{E4}
\operatorname{\hat{X}}=R(\operatorname{X}, \operatorname{Fuzzy} \operatorname{En}(m,r,N))
\end{equation}
{where R denotes the ranking operation and $\operatorname{Fuzzy} \operatorname{En}$ implies the fuzzy entropy matrix of six sensor readings at the current DisPad position. In that way, while the sensor displacement occurs, it is always possible to allow the input of the six sensors to be arranged in order of how strong they represent the pattern of motion, thus reducing the effect of sensor displacement.}
\rev{Note that the ranking operation is performed once at one DisPad position. Besides, to more fairly capture every sensor value change and reduce the impact of noise, we performed max-min normalization on the data and removed the numeric outliers \cite{santoyo2017brief} before entropy calculation.} %
\subsubsection{Refine the Pre-trained Model}
\cg{We employ the transfer learning to generalize the model to other datasets %
with different moving patterns. 
\paragraph{Training Procedure}We use the same parameter settings for both parameters universally %
$\mathcal{D}_{SM}$ and $\mathcal{D}_{MM}$. 
These parameter settings are shown in the appendix.
Then for the above user transferring and motion transferring cases, %
we leverage the same LSTM model structure which is trained on $\mathcal{D}_{SS}$. }
\paragraph{Loss Definition} %
\cg{Same as \cite{zhu2021robust}, we adopted a multi-kernel maximum mean discrepancy (multi-kernel MMD) including over five different Gaussian kernels to measure the distance between the source domain and target domain \cite{zhu2021robust}. MMD maps the distributions {between the source domain and target domain} to a %
reproducing kernel Hilbert space, which returns the distance between the two distributions \cite{gretton2012kernel,gretton2012optimal}. To better estimate the distribution discrepancy, we conduct domain adaptation at the low-dimensional output level \cite{zhu2021robust}. As a result, our MMD loss can be represented as \cite{gretton2012kernel}%
\begin{equation}
    \begin{aligned}
L_{\mathrm{mmd}}=&\left\|\mathbb{E}\left[\psi\left(\hat{Y}_{s}\right)\right]-\mathbb{E}\left[\psi\left(\hat{Y}_{t}\right)\right]\right\|_{\mathcal{H}}^{2} \\
=& \frac{1}{m^{2}} \sum_{i, j=1}^{m} k\left(\hat{y}_{i}^{s}, \hat{y}_{j}^{s}\right)-\frac{2}{m n} \sum_{i, j=1}^{m, n} k\left(\hat{y}_{i}^{s}, \hat{y}_{j}^{t}\right) \\
&+\frac{1}{n^{2}} \sum_{i, j=1}^{n} k\left(\hat{y}_{i}^{t}, \hat{y}_{j}^{t}\right),
\end{aligned}
\end{equation}
where \rev{{$\hat{Y}_{s}$}}
and $\hat{Y}_{t}$ denote the estimated values of the source domain and target domain, respectively.
$\hat{y}_{i}^{s} \in \hat{Y}_{s}(i=1,2, \ldots, m)$ and $\hat{y}_{i}^{t} \in \hat{Y}_{t}(i=1,2, \ldots, n)$ are the respective %
samples of \rev{{$\hat{Y}_{s}$}} 
and $\hat{Y}_{t}$. {$\psi(Y)$} denotes the feature embedding of $Y$ in a reproducing Kernel Hilbert space $\mathcal{H}$ and $k$ is the Gaussian kernel %
function \cite{gretton2012optimal}.
The $L_{\mathrm{mse}}$ is a standard loss of $\mathcal{D}_{SS}$. $L_{\mathrm{mmd}}$ is the loss between the output level's source domain and target domain. Then, the overall loss is denoted as:
\begin{equation}
    L=L_{\mathrm{mse}}+\eta \cdot \lambda \cdot L_{\mathrm{mmd}}, 
    \label{equ:totalloss}
\end{equation}
where $\eta$ is a weighting parameter, and $\lambda$ is a learning rate decaying parameter changing with the iteration of the neural network, which %
prioritizes the minimization of $L_{\mathrm{mse}}$ in the early stage of the training \cite{zhu2021robust}:
\begin{equation}
    \lambda=\frac{2}{1+e^{\frac{-10 \times i}{m}}}-1
    \label{equ:lambda1}
\end{equation}
\begin{equation}
    \lambda=\frac{2}{1+e^{\frac{-10 \times i}{m/10}}}-1
    \label{equ:lambda2}
\end{equation}
where m is a hyperparameter. Due to our small amount of $\mathcal{D}_{SM-TR}$ and $\mathcal{D}_{MM-TR}$ (in our experiment is 2000 ), our model converges very fast. Thus, in the first 5 epochs, we adopted Equation~\ref{equ:lambda1}. Then we used Equation~\ref{equ:lambda2} in the following epochs. We have listed the values \rev{of m and $\lambda$} in the appendix.}
\paragraph{Select Source data for training}Thanks to the ranking operation based on entropy, we make the sensor data in various positions more consistent. That means we do not have to treat all the data from $\mathcal{D}_{SS}$ as the source domain because the variance of different displacements has been reduced. However, to promote the generalization ability of the network, once \rev{the first two entropy ranking indices of $\mathcal{D}_{SS}$ at current DisPad position intersects with that of $\mathcal{D}_{NU}$/$\mathcal{D}_{NM}$}
\begin{figure}[b]
\subfloat[]{\raisebox{0ex}%
{\includegraphics[width=1.8in]{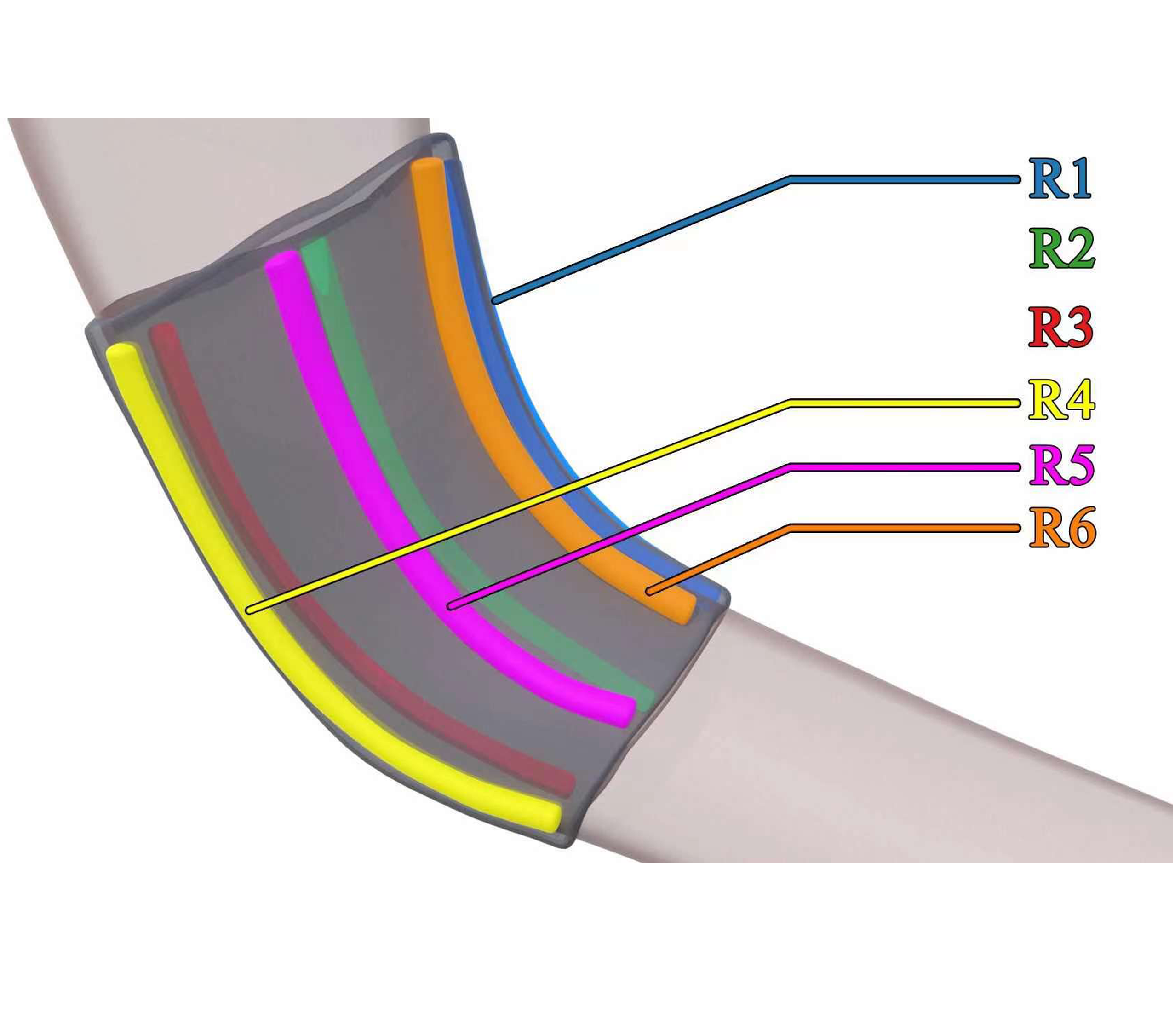}}%
\label{fig:armsingle}}
\hfil
\subfloat[]{\includegraphics[width=1.8in]{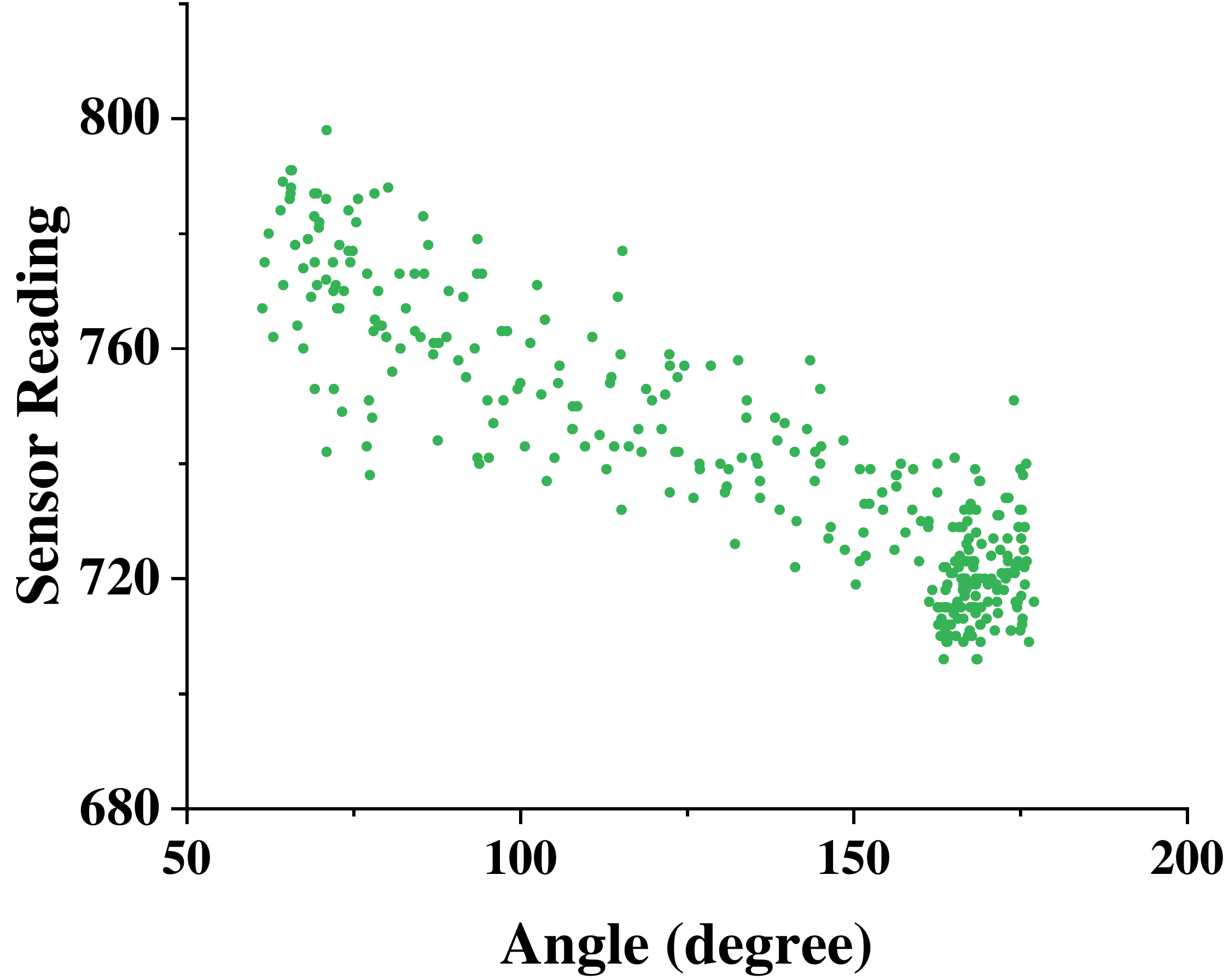}%
\label{fig:singlesensor}}
\hfil
\subfloat[]{\raisebox{0ex}%
{\includegraphics[width=1.8in]{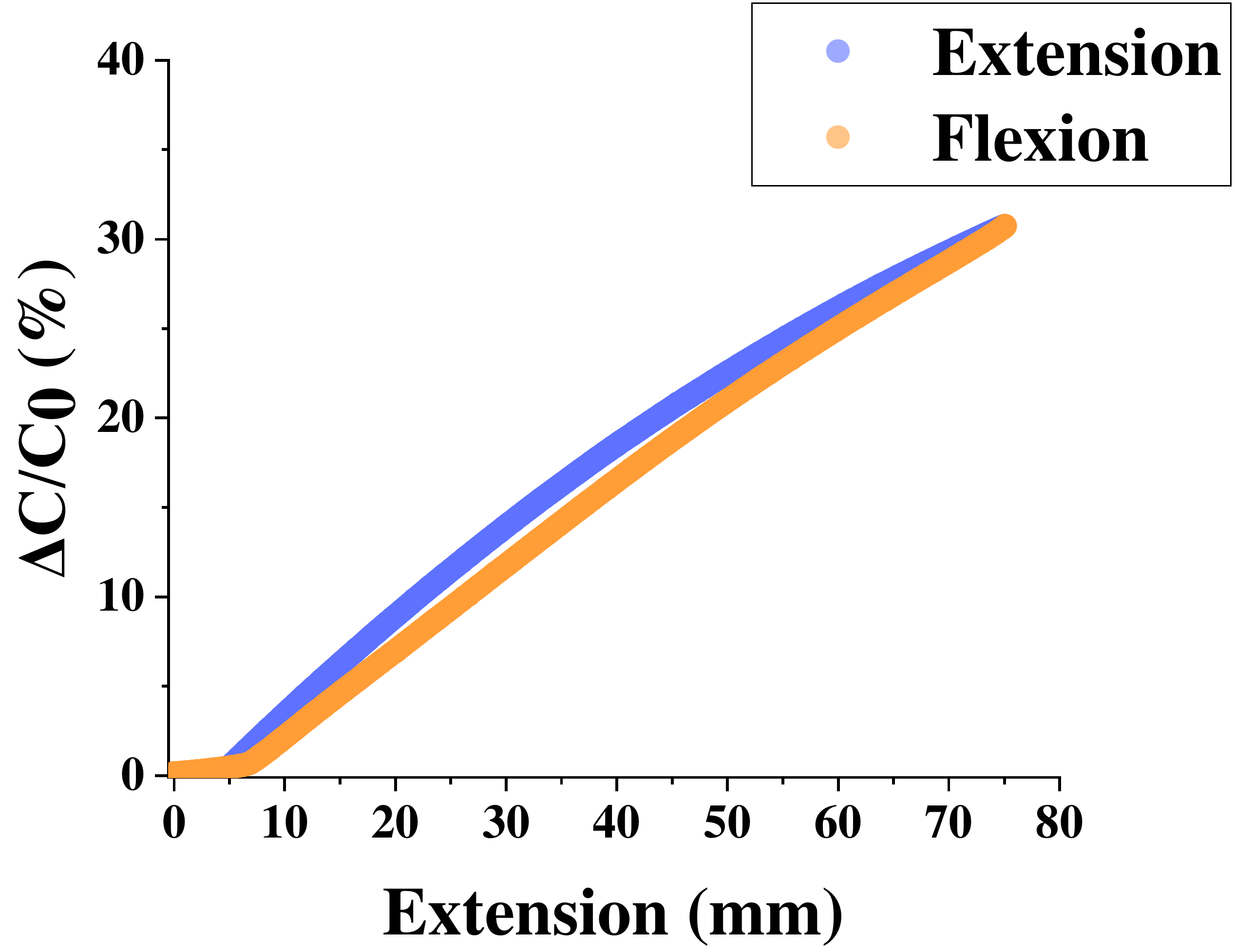}}%
\label{fig:gf}}
\caption{(a) A placement where R2 is on the side of the olecranon of the elbow. (b) The sensor readings correspond to position (a) during the flexion and extension of the elbow. (c) The sensor readings (indicated as R in the vertical) vary with their extension.}
\end{figure}
\section{Implementation}
\subsection{System Prototype}

We have augmented a standard elbow pad with six soft stretchable sensors, which are evenly distributed in a circular pattern around the elbow (Fig.~\ref{fig:elbowpad} and Fig.~\ref{fig:armsingle}). The direction of the arm and the sleeve are parallel, and it can only be worn from the direction as the arrow pointed in Fig.~\ref{fig:elbowpad}.
A further experiment in the result section discusses the intuition and justification of the number of sensors when building the prototype.
These fabric sensors are capacitive and can be bought as off-shelf products from ElasTech \footnote{\url{http://www.elas-tech.com/}}. \cg{The product manual indicates that the sensor can be hand-washed over 100 times and machine-washed over 60 times.}

Fig.~\ref{fig:singlesensor} illustrates a single sensor's reading changing during regular use.
It can be observed that sensor readings vary with different sensor displacements. Fig.~\ref{fig:complexR} depicts the sensor readings over different sensor placements. 
The result shows that for a specific device displacement, sensors on the side of the olecranon are being stretched and produce a larger capacitance as the elbow bends (the angle decreases), creating a linear mapping between the sensor reading and elbow joint angle.
In contrast, other sensors demonstrate more complex and chaotic patterns.

The sensor readings are wirelessly transmitted via Bluetooth Low Energy at a frame rate of \cg{50Hz}. 
The capacitance value is converted to a digital form within [0, 1023].
The circuit board is designed by the sensor vendor \emph{ElasTech} and sends digitalized sensor signals to a mobile phone (Huawei Mate 20, memory: 6GB, Android version: 9) through Bluetooth.
The data are then further transmitted to a remote server, which conducts the training or loading process of the LSTM model.
The circuit can send data directly to a desktop with a Bluetooth dongle.
The server is configured with a one-core CPU without GPU support and a memory capacity of 2GB. The operating system is 64-bit Ubuntu 16.04.
\begin{figure}[t]
  \centering
  \includegraphics[width=0.9\linewidth]{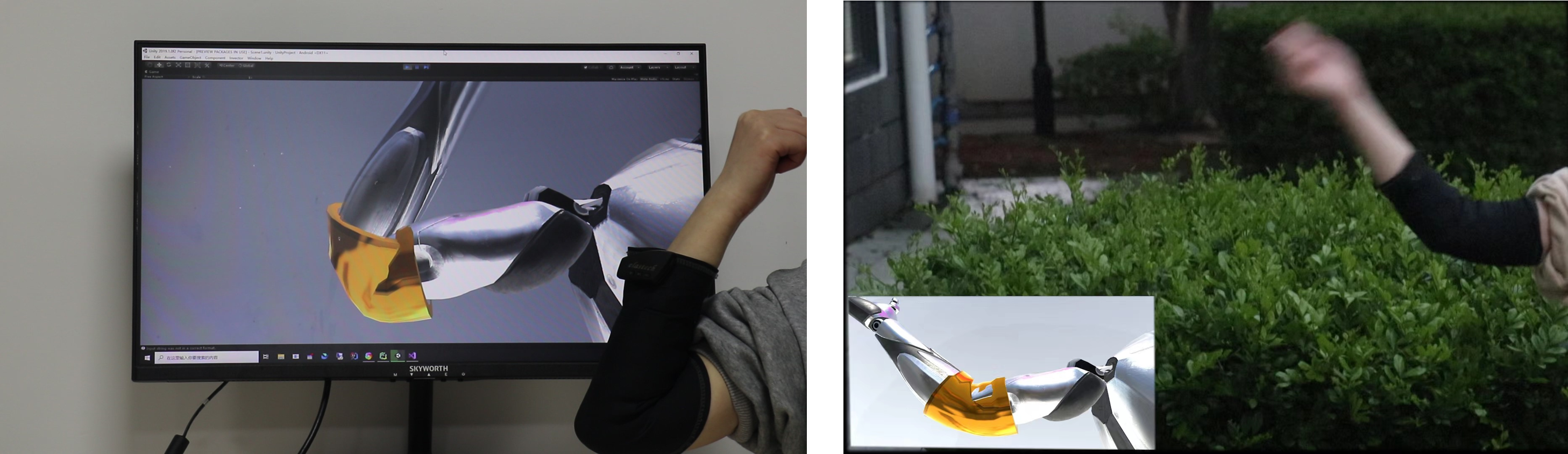}
    \caption{Visualized human joint motion with a computer monitor (left) and a mobile phone (right). }
  \label{fig:visualize}
\end{figure}

\subsection{Dataset Preparation}
The dataset preparation procedure can be divided into three stages: 1) a single subject conducts elbow bending with different device placements. The collected dataset $\mathcal{D}_{SS}$ serves for training and testing the learning model in the standard case; 2) a single subject conducts four motion types (run, walk, jump, clap) to validate our method across different motion types. The collected dataset is named $\mathcal{D}_{SM}$; and 3) a group of participants is recruited to validate our method through a multi-subject experiment, with the collected dataset named $\mathcal{D}_{MM}$. The whole procedure was approved by the Medical Ethis Committee of Xiamen University. Before the data collection, we obtained the participants' written consent after informing them of the experiment's purpose and procedure. 
Here, we define the displacement of our pad along the lateral and circular dimensions as $\eta$ and $\beta$, respectively. 
\begin{figure}[t]
    \centering
    \includegraphics[width=0.9\linewidth]{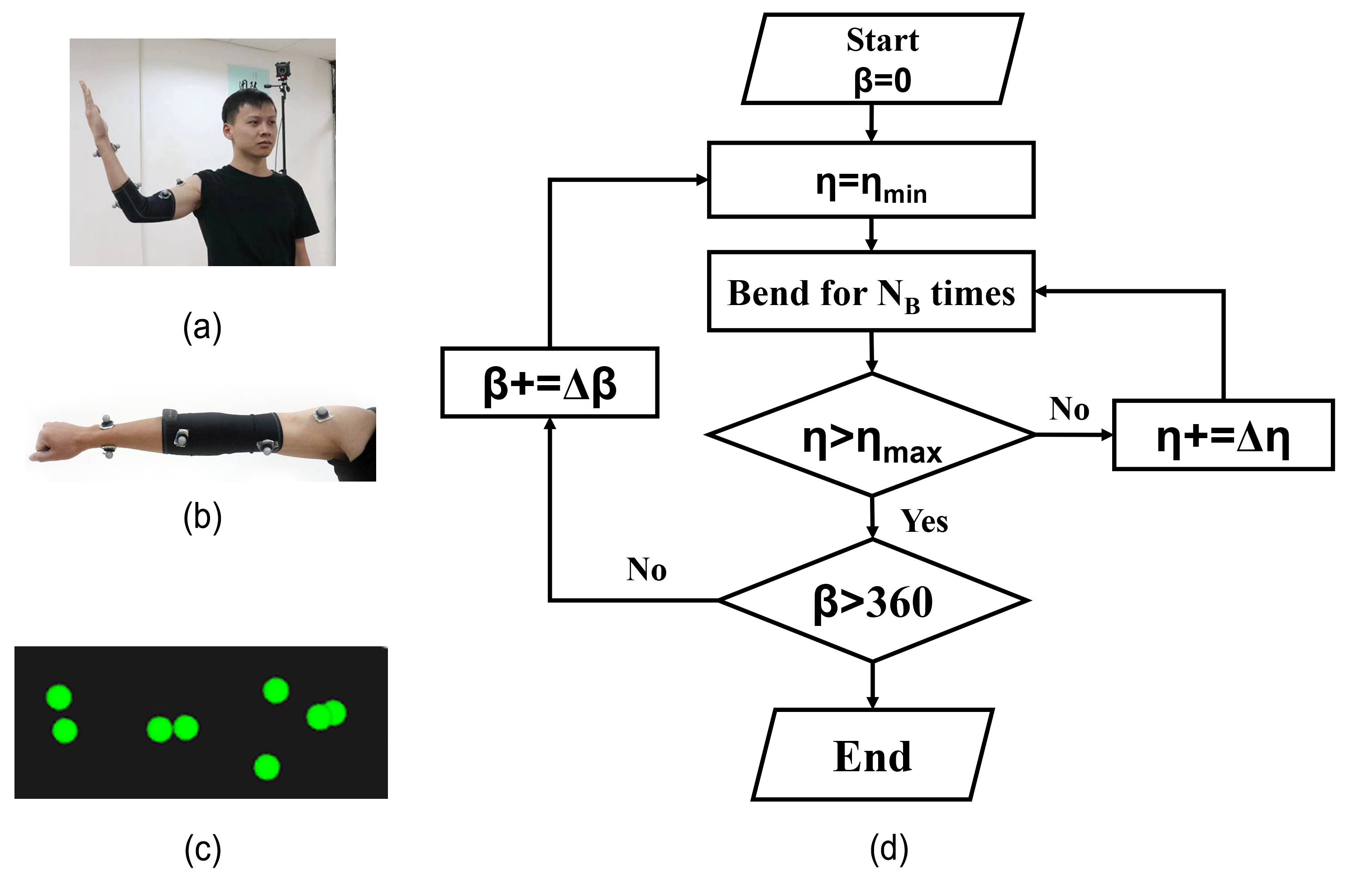}
      \caption{(a) Our subject in a motion tracking studio. (b) DisPad with reflective markers. (c) The markers are captured by the tracking software. (d) The process of collecting training data.}
    \label{fig:CollectDataProcess}
\end{figure}

\subsubsection{Single-subject Single-motion Dataset $\mathcal{D}_{SS}$ Preparation} 
 
\paragraph{Participant}
We invited a participant (Male, Age: 31) to prepare the training dataset. 
The participant (denoted as $\mathbf{P1}$ in the following text) is an engineer at the authors' institution.
The participant's arm girth was 24.75 centimeters.
We conducted a pilot test to evaluate the pad's wearing experience and the sensors' extension (to keep the elbow pad close to the arm such that the sensor could extend effectively). The result shows that the expected arm girth is 20.5 to 28 centimeters.
This ensures a comfortable wearing experience during the process of data collection.

\paragraph{Procedure}
The procedure is approved by the Medical Ethis Committee of Xiamen University.
The participant $\mathbf{P1}$ entered a motion tracking studio wearing the pad (Fig.~\ref{fig:CollectDataProcess}a).
The pad was augmented with reflective markers that were tracked by the infrared cameras (Fig.~\ref{fig:CollectDataProcess}b).
The 3D positions of the markers were tracked and computed by the Nokov motion-capture system at a rate of 60 FPS (Fig.~\ref{fig:CollectDataProcess}c).
From the tracked positions in Fig.~\ref{fig:CollectDataProcess}c, we can calculate the vector between the centerline of the upper arm and the lower arm, and the angle between these two vectors is the rotation of the elbow joint.
We synchronized the collected signals from the sensors and the joint angles recorded by the motion-capture system by triggering the two software programs concurrently. 
Given the different frame rates of these two signal sources, we down-sampled the joint values to ensure the consistency of the sample size.
These routines were applied to the following data preparation procedures throughout the paper.

During the collection of single-subject single-motion dataset $\mathcal{D}_{SS}$, the participant's action sequence was designed as Fig.~\ref{fig:CollectDataProcess}d. 
The range of the circular displacement was $\beta\in$[0$^\circ$, 360$^\circ$], which indicates that we consider various flexibility of allowing users to rotate the pad largely unrestricted.
We defined the chelidon as our starting position though the starting position can be any point in the circle around the junction of the upper and lower arms. 
The range of the lateral displacement was defined as $\eta\in$[$\eta_{min}$, $\eta_{max}$], where $\eta_{max}$=-$\eta_{min}$=4 cm. 
For each trial, the participant conducted $N_B$=8 bending cycles.
For each bending cycle, the participant bent the elbow within the range of rotation (approximately $\theta\in$[40$^\circ$, 180$^\circ$]) at different velocities.
After each bending trial, we shifted the pad by a slight lateral displacement $\Delta \eta$=1 cm until it reached $\eta_{max}$. After an entire cycle of lateral displacement, the pad rotated by $\Delta \beta=5^{\circ}$, and the previous procedures were repeated.
The collection took around 1.2 hours in total.

\subsubsection{Single-subject Multi-motion Dataset $\mathcal{D}_{SM}$ Preparation}
\paragraph{Participant}
The same participant is involved in the procedure of Single-Subject Single-Motion dataset preparation.

\paragraph{Procedure} 
During the collection of the motion dataset $\mathcal{D}_{SM}$, the participant was instructed to perform four daily activities: walking, running, clapping hands, and jumping.
Since people perform the selected activities
frequently in their daily activities, the data collection procedure requires no further pre-training. They can be used as a reference for our prototype application in daily routines. 
Walking, running, and jumping are routine everyday activities that exhibit a range of magnitude, resulting in different rotation speeds and ranges of the elbow joint.
Within each trial, each type of activity lasted for one hundred seconds.
We conducted the trial three times after shifting the pad by a random circular and lateral displacement during each time.
The displacement configuration was within the range ($\beta\in$[0$^{\circ}$, 360$^{\circ}$], $\eta \in$[$\eta_{min}, \eta_{max}$]). The whole collection process lasted 22 minutes.

\subsubsection{Multi-subject Multi-motion Dataset $\mathcal{D}_{MM}$ Preparation}

\paragraph{Participants}
We invited ten subjects to participate, including five males $\mathbf{P4, P8-11}$ %
and five females $\mathbf{P2, P3, P5-7}$. Their ages ranged from 21 to 55, and their arm girths were between 20.5 centimeters and 28 centimeters.
They were all graduate students in the authors' institution.

\paragraph{Procedure}
To construct the dataset $\mathcal{D}_{MM}$, each participant followed the procedure of collecting $\mathcal{D}_{SM}$ except that they collected data at each random displacement for 40 seconds, respectively. The arrangement of markers is the same as Fig.~\ref{fig:CollectDataProcess}b and Fig.~\ref{fig:CollectDataProcess}c.
The collection process lasted 21 minutes in total.
After the collection, we did an interview to figure out the wearing experience.

\subsection{Network and Training Details}
\subsubsection{Network Model}
 \begin{table}[t]
      \caption{Statistics and partition of datasets, in terms of DisPad positions.}~\label{tab:dataset}
    \centering
    \begin{tabular}{ccccc}
       \toprule 
	   & Training & Validating & Testing & Total \\
 		\midrule
 		$\mathcal{D}_{SS}$  & 378 & 126 & 135 & 639\\
 		\midrule 
       $\mathcal{D}_{SM}$& 4 & 4 & 4 & 12\\
       \midrule 
       $\mathcal{D}_{MM}$& 10 & 10 & 10 & 30 \\
       \bottomrule
    \end{tabular}
\end{table}

\cg{We \rev{considered} the problem of estimating the angle of an elbow joint %
as a regression problem. To address this problem, 
we {constructed} a six-layer %
LSTM model \cite{hochreiter1997long} to match the sensor readings and optical ground truth. 
The network input \rev{was} 30 frames of sensor readings (size 30x6), and the output \rev{was} the elbow joint angle (size 1x1). The batch size \rev{was} 256, and the hidden units \rev{were} 256, too. The learning rate \rev{was} 0.01 and reduces by {10\%} every two epochs. To avoid gradient explosion and gradient extinction, we adopted%
the gradient clip.
A batch normalization layer was added to unify the data distribution for each layer. In the output layer, a fully connected layer \rev{transformed} the output of the LSTM layer to the estimated elbow joint angle. \rev{To make the output smoother, we adopt an error-state Kalman filter. The ratio of predicted error to measure error is 2.67.}

\subsubsection{Loss Definition for Angle Prediction}
The MSE loss is calculated between the sensor readings and optical ground truth of $\mathcal{D}_{SS}$: 
\begin{equation}
    L_{\mathrm{mse}}=\frac{1}{n} \sum_{i=1}^{n}\left(\hat{y}_{i}-y_{i}\right)^{2}, 
\end{equation}
where $\hat{y}_{i}$ denotes the estimated value of the LSTM model, $y_{i}$ denotes the optical ground truth.
$n$ is the number of samples in a batch.
\subsubsection{Data Statistics \& Partition}
\cg{The final $\mathcal{D}_{SS}, \mathcal{D}_{MM}$, and $\mathcal{D}_{MM}$ datasets contain 219,932, 60,276, and 60,972 samples, respectively. The datasets and the code will be released to the public after the publication of this paper.
Since we observe an excessive number of samples in the circular displacement range of $\beta$=[160, 180] in the original training dataset, we discussed the impact in Section \ref{sec:trackingondss}.
\paragraph{Training for Single User} When parting the $\mathcal{D}_{SS}$ dataset, the DisPad positions are randomly divided into three %
partitions (training, validating, testing) so that the samples in the three partitions are independent of each other.

\begin{figure}[b]
\subfloat[]{\raisebox{0ex}%
{\includegraphics[width=2.5in]{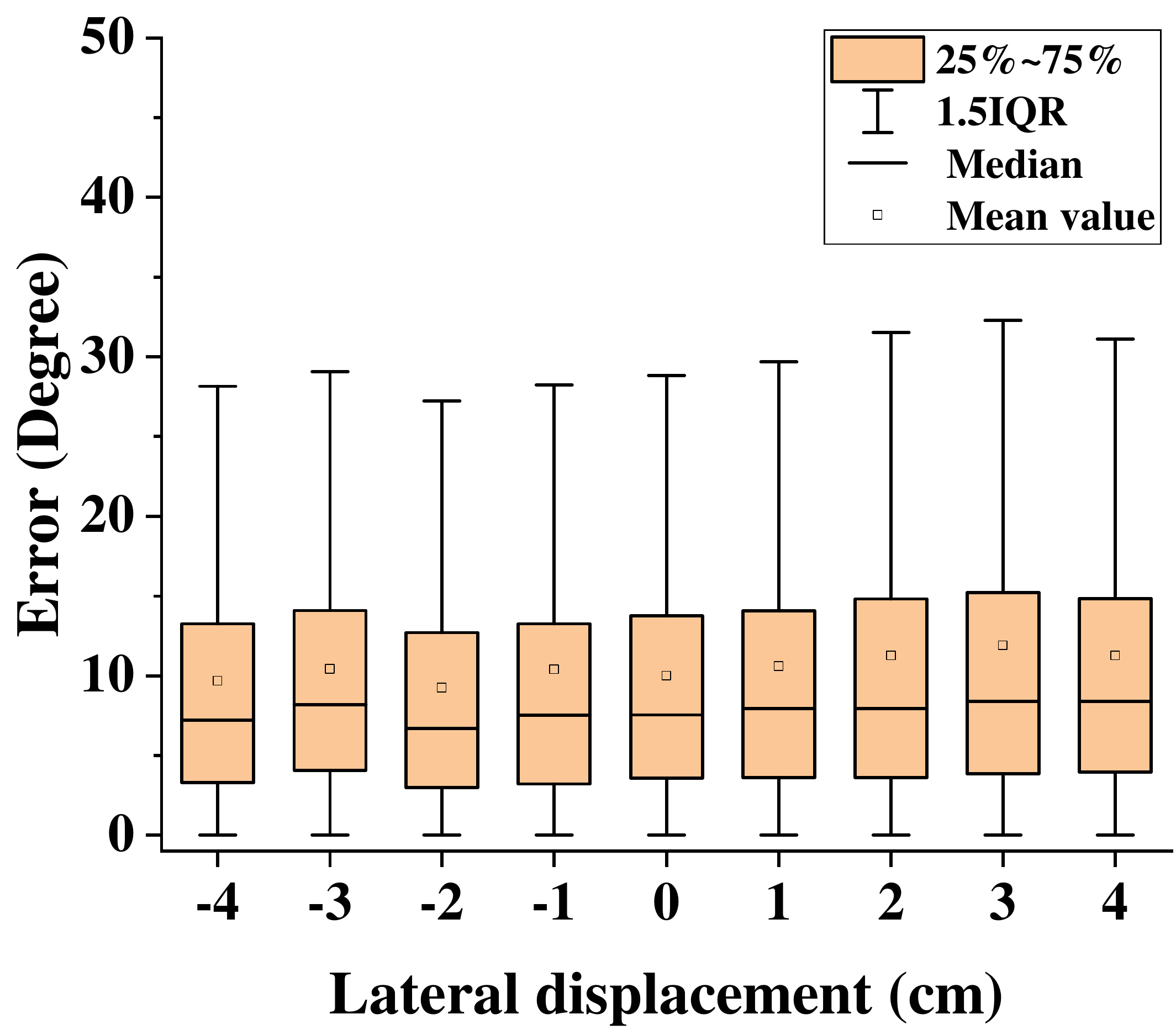}}%
\label{fig:lateral}}
\hfil
\subfloat[]{\includegraphics[width=3.2in]{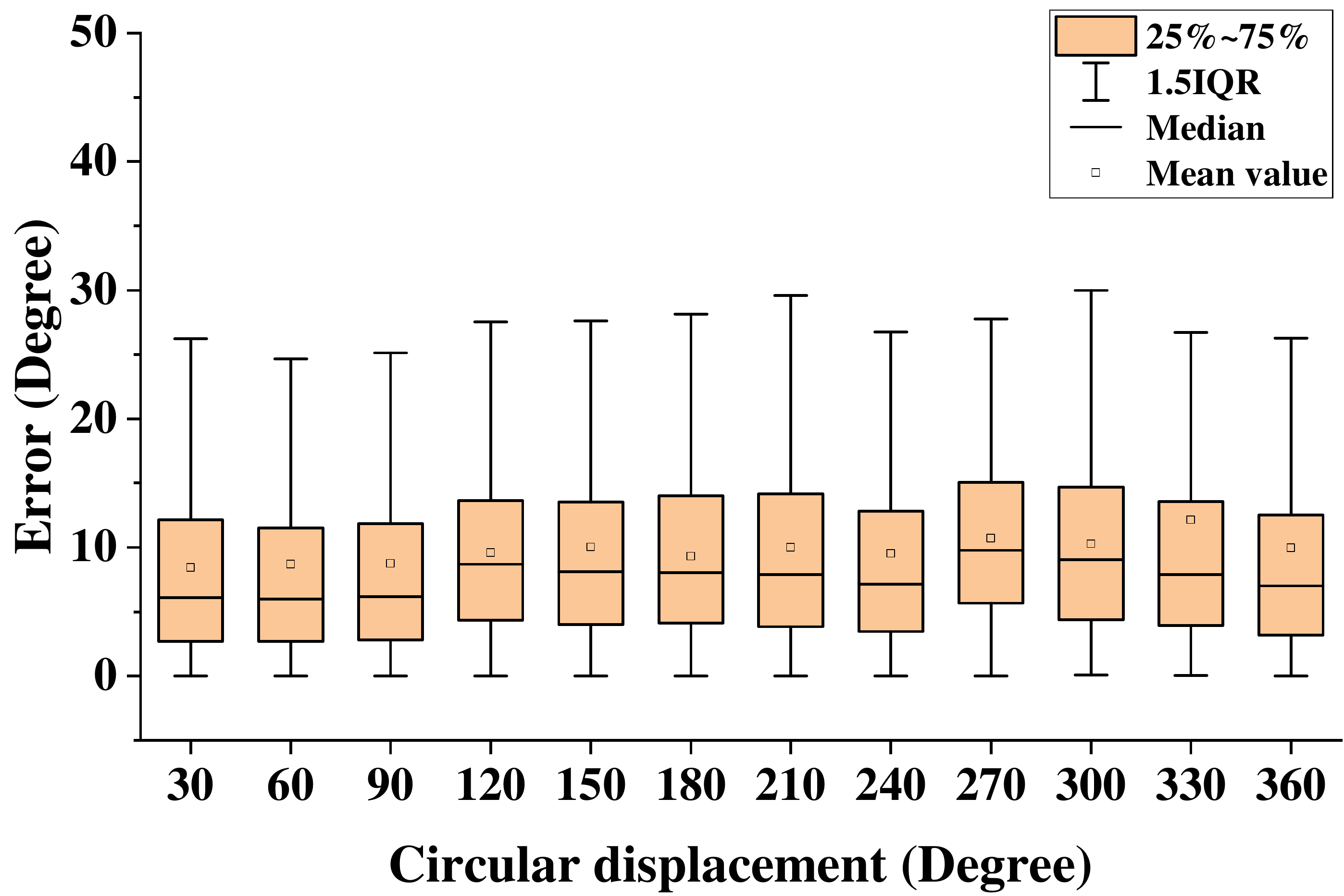}%
\label{fig:circular}}
\caption{(a) The average tracking errors over individual lateral displacements $\eta$. (b) The average tracking errors over individual circular displacements $\beta$.}
\label{fig_sim1}
\end{figure}

Table~\ref{tab:dataset} lists the dataset statistics and their partition into training, validation, and testing.
 \paragraph{Source Domain for Transfer Learning}
 \rev{This selection can refer to $\mathcal{D}_Source Train$ in Alg.~\ref{alg::ag1}}.
 \paragraph{Target Domain for Transfer Learning}Since we meant to transfer the model to new motions (or users) with a simple %
 data acquisition effort, we only adopt 2,000 samples of one DisPad position of target motion (or user) for training and validate the result on the remaining DisPad positions. Note that the labels of all training sets are removed.

 This partition ensures that the trained model does not over-fit in the training partition and can be generalized to new DisPad positions in the testing dataset.

The value in each cell indicates the number of DisPad positions in each condition.
In the following discussions, we use the suffix ($-TR, -VA, -TE$) to indicate the partition (e.g., $\mathcal{D}_{SS-TR}$ denotes the training subset of single-subject single-motion dataset  $\mathcal{D}_{SS}$).}}

\section{Result}

\subsection{Tracking Results on Single-subject Single-motion Dataset $D_{SS}$}
\label{sec:trackingondss}
\begin{figure}[htbp]
\subfloat[]{\raisebox{0ex}%
{\includegraphics[width=2.4in]{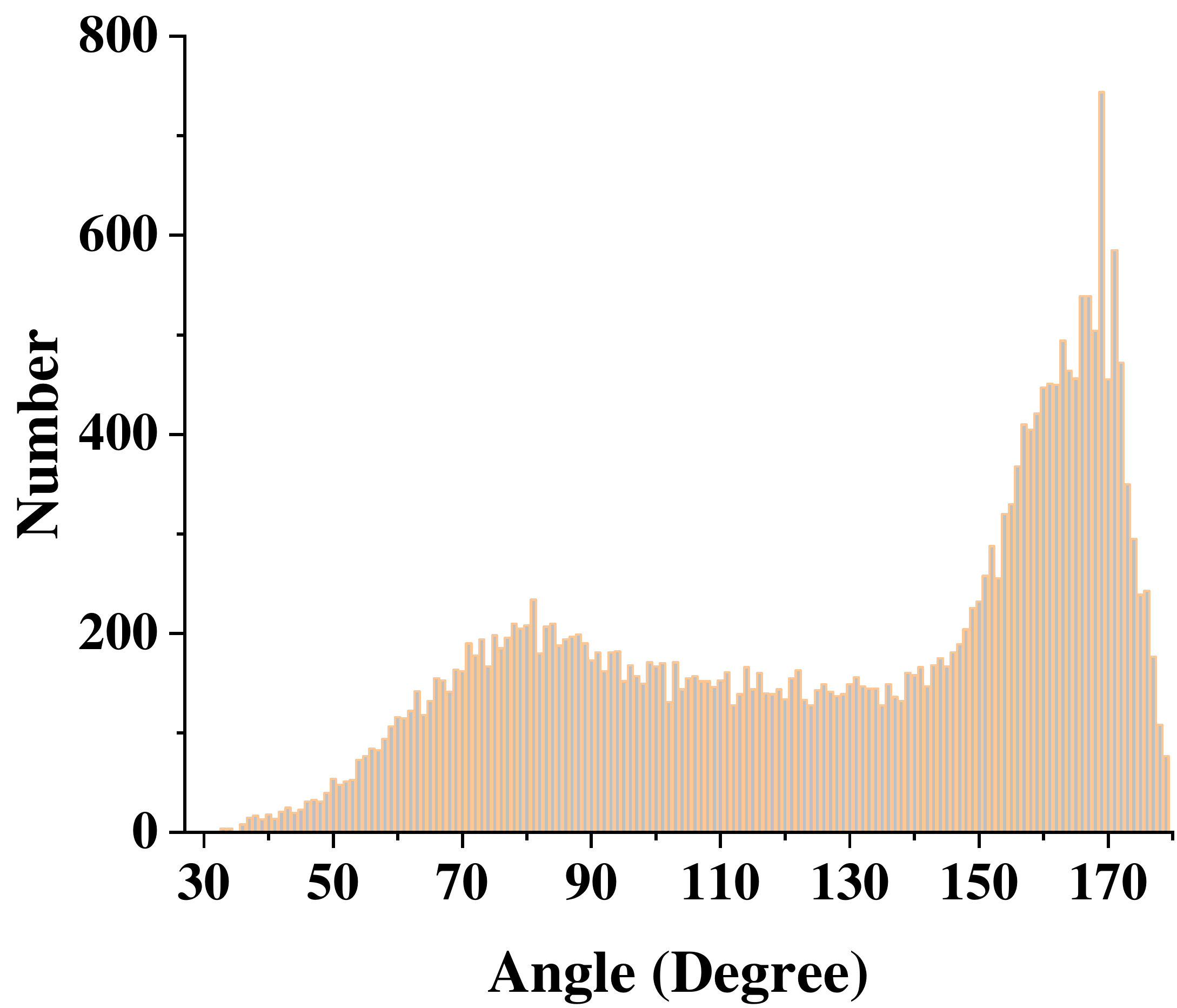}}%
\label{fig:datadistri}}
\hfil
\subfloat[]{\includegraphics[width=2.4in]{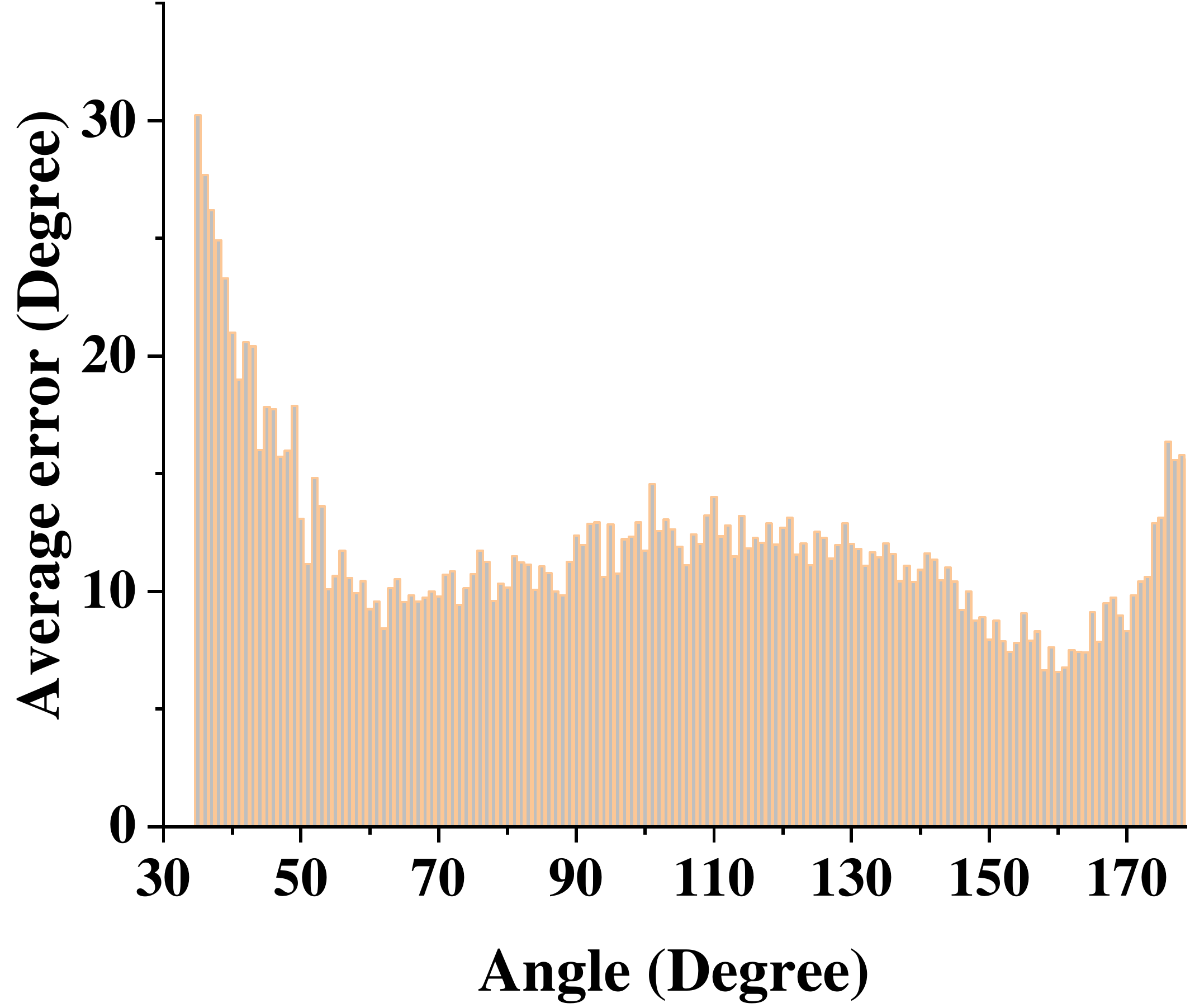}%
\label{fig:angleaccu}}
\hfil
\subfloat[]{\includegraphics[width=2.4in]{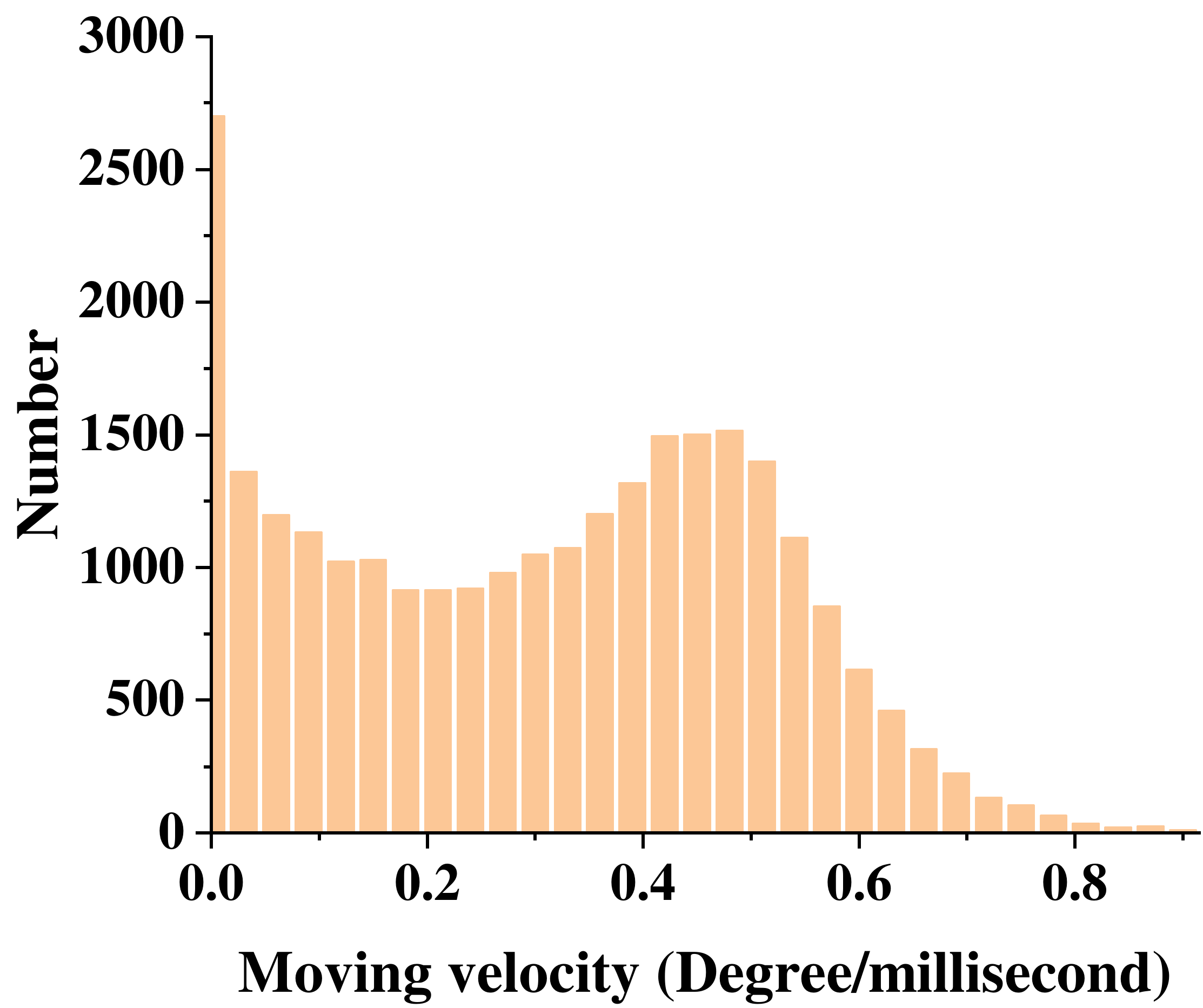}%
\label{fig:velnumber}}
\hfil
\subfloat[]{\includegraphics[width=2.4in]{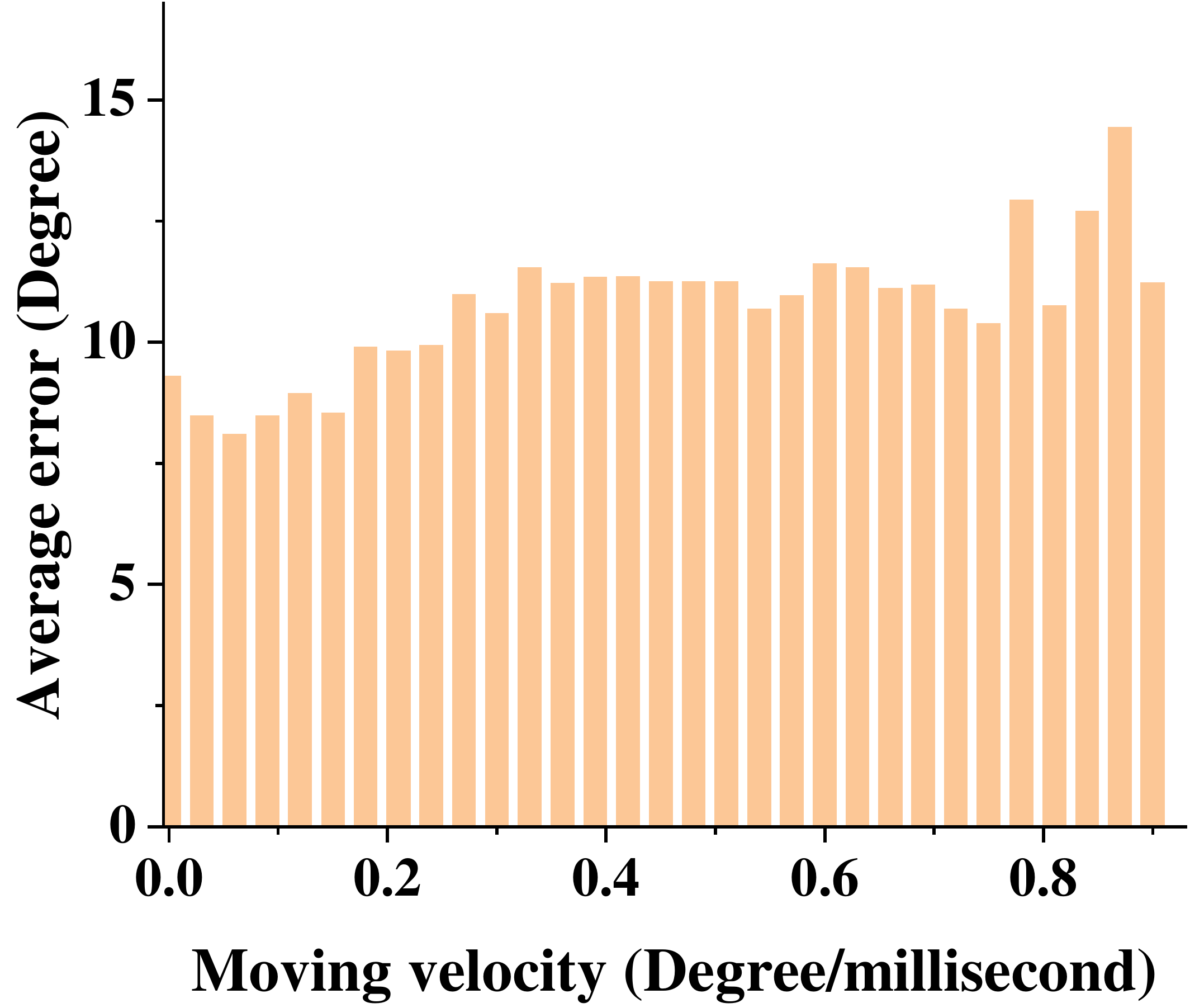}%
\label{fig:vel}}

\caption{(a) The relationship between the number of data points and bending angles. (b) The tracking errors of the method under different bending angles. (c) The relationship between the number of data points and different moving velocities. (d) The tracking errors of the method under different moving velocities.}
\label{fig:datavel}
\end{figure}

During the training stage, we optimize the learning-based model on the $\mathcal{D}_{SS-VA}$ and compute an average of tracking errors on the $\mathcal{D}_{SS-TE}$.
After 98 seconds of training, the final average tracking error of estimating the bending angle of the elbow joint in $\mathcal{D}_{SS-TE}$ is 9.82 degrees.

The tracking errors with respect to the lateral displacements and the circular displacements are illustrated
in Fig.~\ref{fig:lateral} and Fig.~\ref{fig:circular}, which imply that the lateral and circular displacements have no apparent effect on the tracking errors. This verifies that our method can be applied to various lateral and circular displacements, indicating the robustness of our algorithm.

We evaluate our results on the \emph{testing dataset} based on the joint bending angles and moving speed. The impacts of the joint bending angles and moving speed are depicted in Fig.~\ref{fig:angleaccu} and Fig.~\ref{fig:vel}, respectively, while the number of data points for different joint bending angles is illustrated in Fig.~\ref{fig:datadistri}. 
Fig.~\ref{fig:datadistri} shows the data points of the joint bending angle are mainly concentrated in the ranges [160, 180]. 
In contrast, there are fewer data points around 30 degrees. 
From Fig.~\ref{fig:angleaccu}, we find that the maximum and minimum tracking errors are located around the angle values of 30 and 160 degrees, respectively. 
The minimal \rev{errors} might be attributed to a large number of data points around 160 while the maximum \rev{errors} may be due to a small number of data points in the range [30,50].
The correlation value between the number of samples and the average errors at a specific joint angle is \cg{-0.60}. 
This implies that they are moderately negatively correlated, i.e., increasing the number of samples could result in a tendency to reduce the tracking error.
As a consequence, the number of samples at a particular joint angle in the dataset may have an effect on the error. 
Fig.~\ref{fig:vel} depicts the relationship between the velocity and the average tracking errors. The velocity is calculated by the sum of the angle variation divided by the time spent with a time window of 30 ms. In addition, we conducted a correlation analysis to figure out if there is a correlation between moving velocity and average tracking errors. \cg{The result is 0.78}, which implies that there is a highly positive correlation between the moving velocity and the tracking errors. To explore the reason, Fig.~\ref{fig:velnumber} shows the relationship between the number of data points and moving velocity. We can infer from this figure that an excessive number of data points around 0 degrees/ms creates a low average tracking error. On the contrary, the number of data points decreases dramatically while the moving velocity exceeds 0.5 degrees/ms and reaches the minimal value at 0.8 degrees/ms. This trend reverses Fig.~\ref{fig:vel}, which peaks at 0.8. The correlation analysis shows that 
the tracking errors and number of data points are moderately related  (the result is -0.46). Hence, the errors are also correlated to the number of data points.

\subsection{Tracking Results on Single-subject Multi-motion Dataset $D_{SM}$}

\begin{figure}[b]
\subfloat[]{\raisebox{0ex}%
{\includegraphics[width=1.5in]{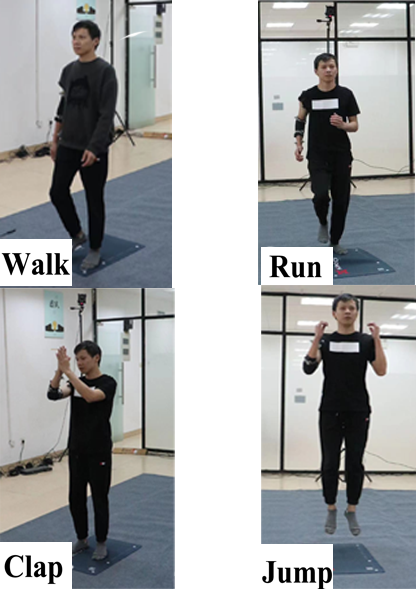}}%
\label{fig:fourmotion}}
\hfil
\subfloat[]{\includegraphics[width=2.7in]{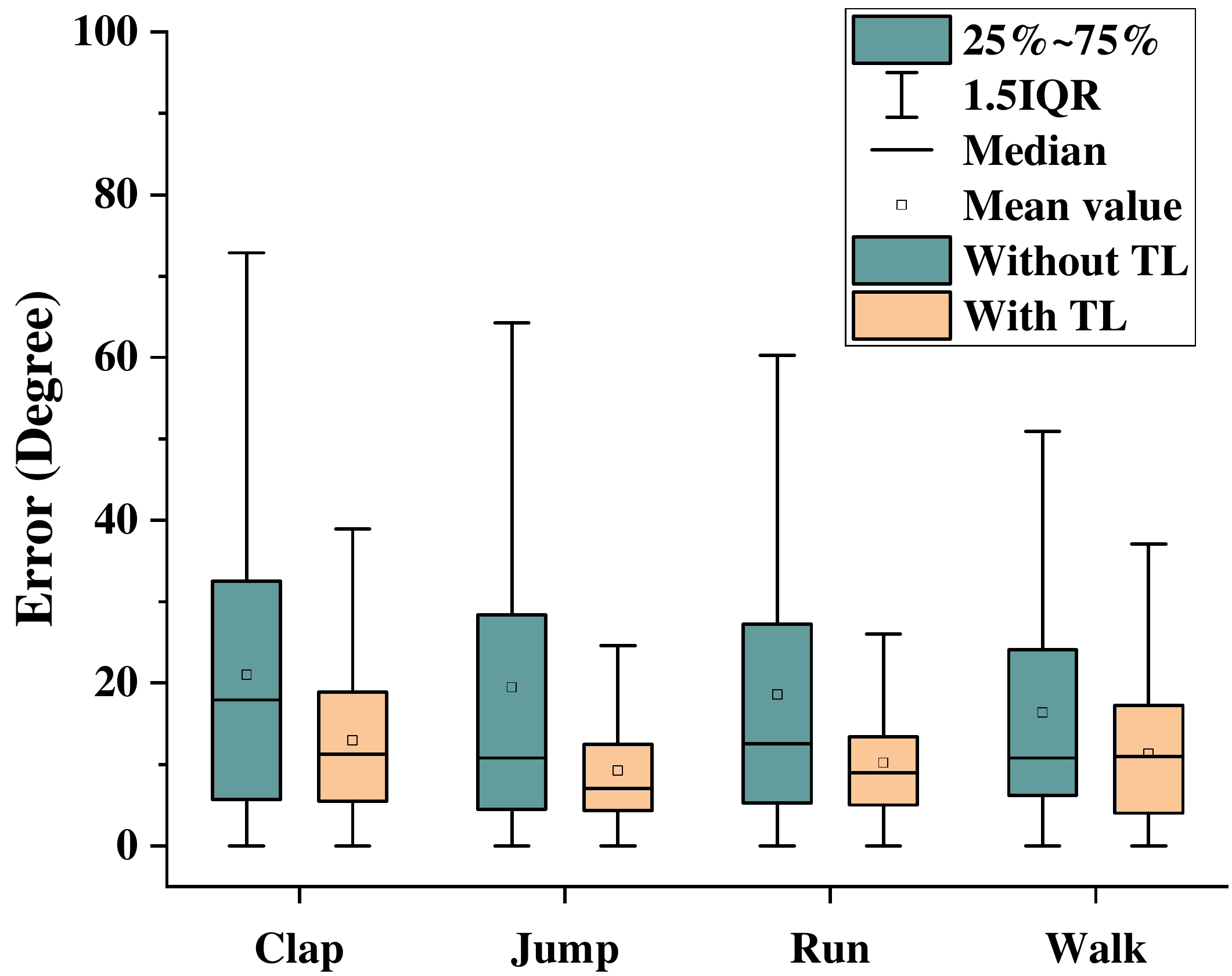}%
\label{fig:motioncompare}}
\caption{ (a) Representative images of the process of building the motion dataset $\mathcal{D}_{SM}$. (b) The average tracking errors of different motion types in the motion dataset $\mathcal{D}_{SM}$ and the effect before and after transfer learning.}
\label{fig:MotionType}
\end{figure}

Representative images of the process of capturing scenes of different motion types are illustrated in Fig.~\ref{fig:fourmotion}. The placement of markers on the elbow is the same as that shown in  Fig.~\ref{fig:CollectDataProcess}b. The relationship between different motion types and the average tracking errors is depicted in Fig.~\ref{fig:motioncompare}. We also compare the tracking errors with or without transfer learning. 
\begin{figure}[b]
\subfloat[]{%
{\includegraphics[width=5in]{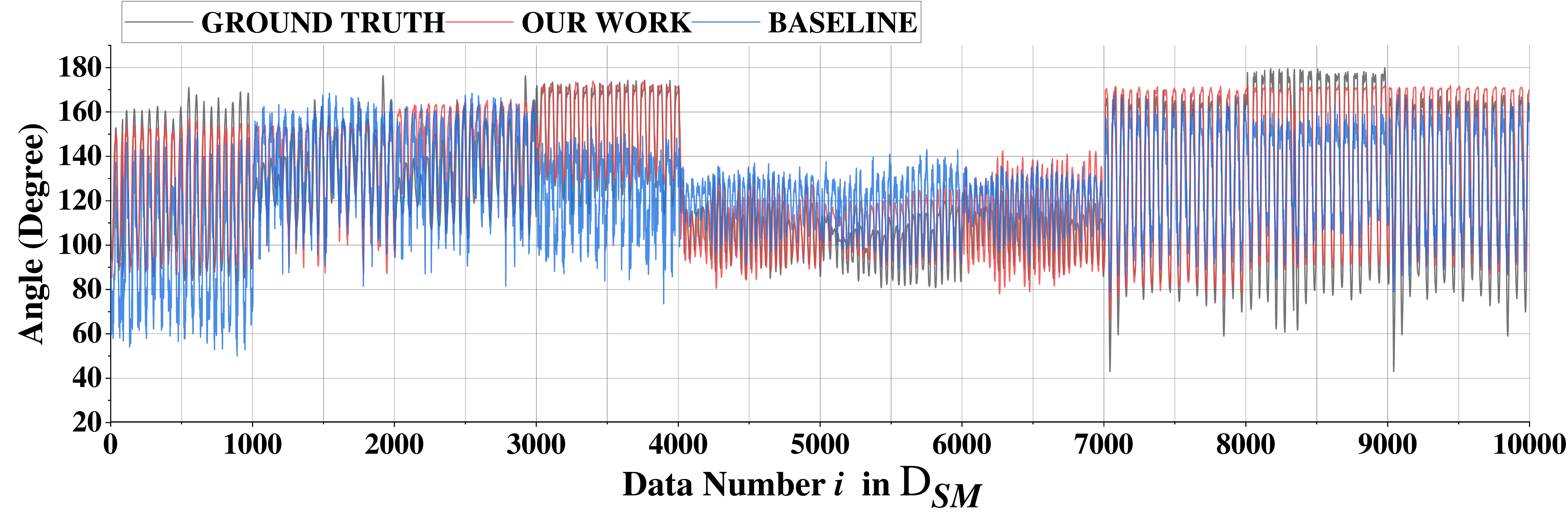}}%
\label{fig:detailmotion}}
\hfil
\subfloat[]{\includegraphics[width=5in]{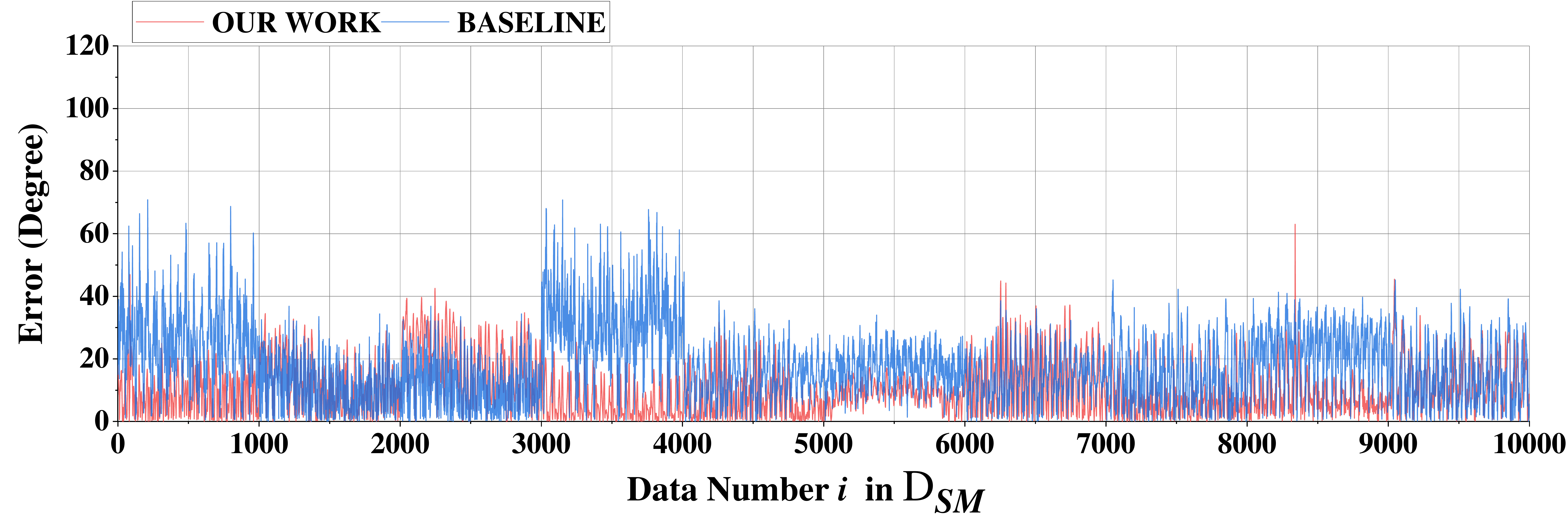}%
\label{fig:detailmotionerr}}
\caption{(a) A plot of the detailed predict results on $\mathcal{D}_{SM}$. (b) A plot of the detailed predict errors in on $\mathcal{D}_{SM}$.}
\end{figure}
\cg{It can be easily seen that without transfer learning, the tracking errors of clapping are the highest on average, followed by jumping, then running and walking. Their average tracking errors surpass %
fifteen degrees. Besides, the difference between the upper limit and the lower limit exceeds 40 degrees.}

\cg{In contrast, %
as shown in Fig.~\ref{fig:MotionType}b, the difference between the four motion types is smaller with transfer learning. In addition, the average tracking error and the difference between the upper limit and the lower limit also decrease. The upper limit is now below 40 degrees while the average tracking is around 10 degrees.
To further measure the effect of the transfer learning, we ran the Mann-Whitney U test (given that the data do not follow a normal distribution (p\textless0.05)) to check if the difference of the tracking errors before and after transfer learning is statistically significant. The result (U=1534615168, p\textless 0.001) indicates the significant difference between the errors before and after transfer learning. It proves that transfer learning is 
suitable for %
the prediction of different motion types and reduces the tracking errors of different motion types. We also verified the average tracking error on the dataset $\mathcal{D}_{SM}$, which is 10.98 degrees.}

\cg{Fig. %
\ref{fig:detailmotion} and  %
\ref{fig:detailmotionerr} %
show our predicted result compared to the ground truth and baseline (without ranking entropy and transfer learning) on %
$\mathcal{D}_{SM}$. It can be observed that there are different motion patterns in Fig.~\ref{fig:detailmotion} and the samples of each motion were similar. Although the baseline prediction results have a roughly similar shape to the ground truth, 
the baseline prediction is not as accurate as our method in moving ranges (Fig.~\ref{fig:detailmotionerr}). This difference means that the data distribution of $\mathcal{D}_{SM}$ is not similar to that of $\mathcal{D}_{SS}$ owing to sensor displacements and new motion patterns. Compared to the baseline, our model generalizes better on those different patterns. %
}
\subsection{Tracking Results on Multi-subject Multi-motion Dataset $D_{MM}$}

\begin{figure}[t]
    \centering
    \includegraphics[width=0.9\linewidth]{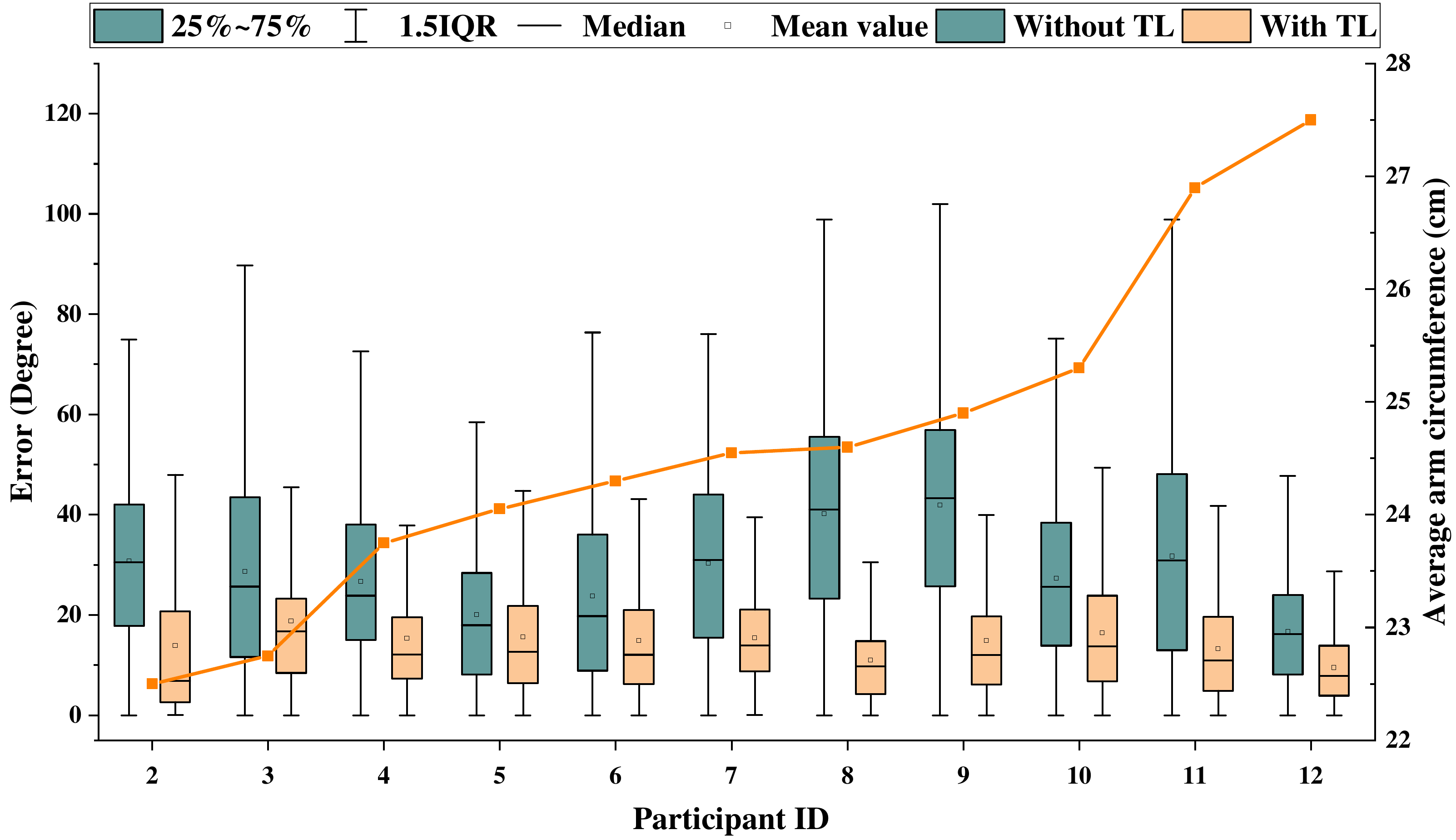}
      \caption{\cg{The tracking errors and the arm \rev{girths} of each participant.}}
    \label{fig:diffUsers}
\end{figure}

\cg{Since we built the training dataset $\mathcal{D}_{SS}$ based on the participant \textbf{P1}, we test the performance on %
the multi-subject multi-motion dataset $\mathcal{D}_{MM}$ to measure the generalization ability of DisPad to %
users. 
As shown in Fig.~\ref{fig:diffUsers}, the tracking errors without transfer learning are %
obviously different among %
the ten users. However, the tracking errors and the range of errors reduce significantly after the transfer learning is applied. %
We ran the Mann-Whitney U test (given that the data do%
not follow the normal distribution (p value\textless0.05)), confirming a significant difference between the errors before and after transfer learning (U=1383015500.00, p\textless 0.001).

With %
transfer learning, the average tracking error on %
the dataset $\mathcal{D}_{MM}$ is 11.81 degrees. This shows the robustness across different users. To explore a possible relationship between the users' arm girth values and the tracking errors, we measure the average arm \rev{girths} (the average value of the upper and lower arm \rev{girths}) of the corresponding users. The data
are also shown in Fig.~\ref{fig:diffUsers}. The %
correlation between the {users'} arm \rev{girths} and the average tracking error with %
transfer learning is %
0.11, indicating a weakly positive correlation. %
This is not surprising because \textbf{P1}'s arm girth is 24.75 {cm}, which is smaller than most of the other users' arm \rev{girths}. Thus, smaller arm girth leads to better prediction results in this case.}

\begin{figure}[t]
\subfloat[]{%
{\includegraphics[width=5in]{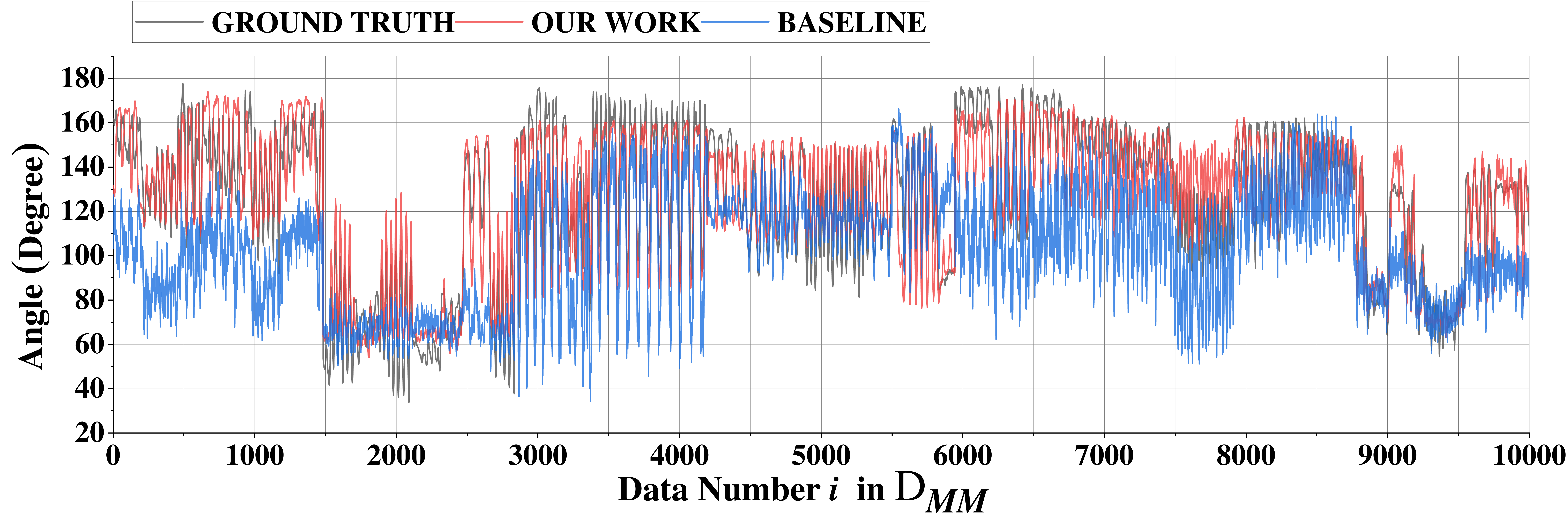}}%
\label{fig:detailuser}}
\hfil
\subfloat[]{\includegraphics[width=5in]{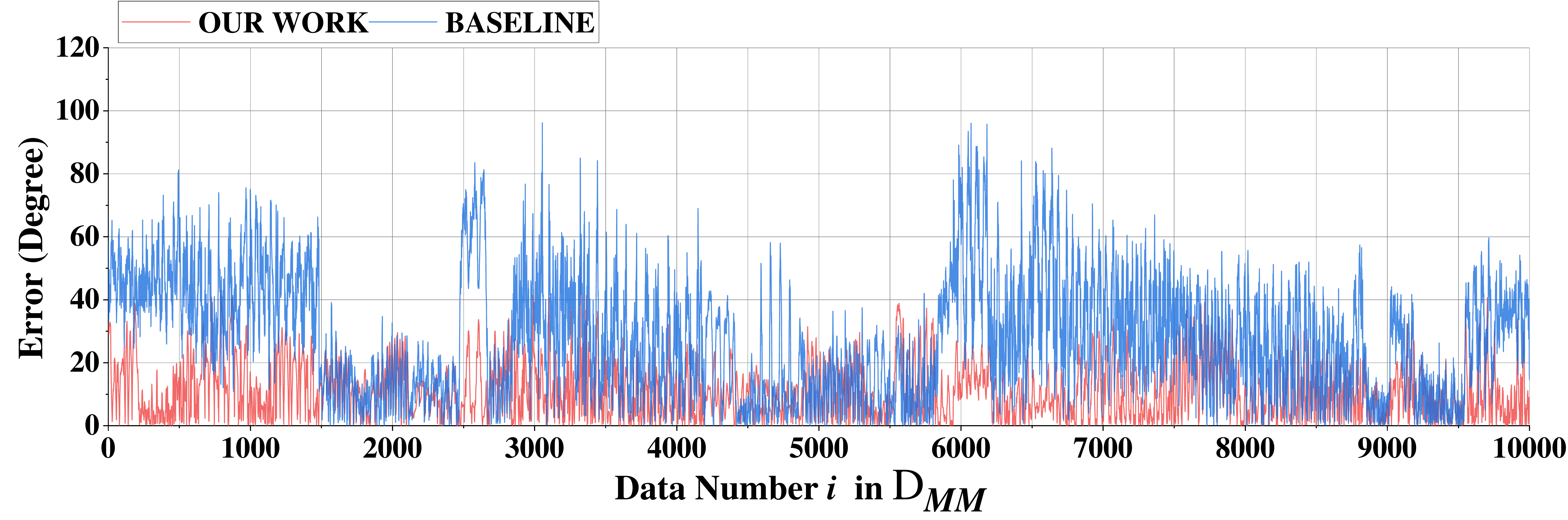}%
\label{fig:detailusrerr}}
\caption{(a) A plot of the detailed predict results on $\mathcal{D}_{MM}$. (b) A plot of the detailed predict errors on $\mathcal{D}_{MM}$.}
\end{figure}

\cg{Fig.~\ref{fig:detailuser} and~\ref{fig:detailusrerr} illustrate the predicted results on $\mathcal{D}_{MM}$. Compared to Fig. %
\ref{fig:detailmotion}, the {moving range and duration of the motion} are more complex due to the random motions every user performed %
in a short time. Attributed to the above reasons, %
the baseline failed to predict accurate moving ranges, causing bigger errors than ours (Fig.~\ref{fig:detailusrerr}). On the contrary, our model outperforms the baseline and generalizes well on %
$\mathcal{D}_{MM}$.}

\subsection{User Feedback on Wearing Experience}

\cg{During the interview after capturing the dataset %
$\mathcal{D}_{MM}$, most of the users mentioned that it was %
comfortable to wear DisPad and it would %
not impede them from doing movements. However, $\mathbf{P11}$ acknowledged that the wearing experience was %
not bad but it was %
inconvenient for moving because of the increasing %
pressure of the DisPad. %
He %
also said that it could be improved by making a DisPad of a bigger size. $\mathbf{P3}$ also supported %
this view: ``It will be better if the size can change with some technologies to %
fit more people with different arm girths.'' 

To validate the tracking effect in new users, we invited nine users (denoted as $\mathbf{P12}$ to $\mathbf{P20}$), whose arm \rev{girths} %
\rev{are} between 20.5cm and 28cm, for evaluation and also for video shooting. The procedure was like this: First, they put on the elbow pad for calibration (the calibration included collecting 2,000 samples of data for ranking and training the transferring model).
Generally, the calibration process only lasted for 2.7 minutes (including the 2,000-sample data-collecting process). Note that every user calibrated only once, and then they could wear the DisPad in various displacements without further calibration. After the calibration, those users were allowed to bend their arms while watching the visual effects. Lastly, we did a video shoot to present the tracking effects. Also, the whole procedure was reviewed and approved by the Medical Ethics Committee of Xiamen University. Before the interview and video shoot, we obtained each participant's written consent after informing them of the experiment's purpose and procedure. 

After the experiment, we interviewed the participants %
to investigate the wearing experience, whether the elbow affected %
the user's motion, and whether the visualized result was %
accurate. During the interview, all the users mentioned that the elbow pad was %
comfortable and it did %
not affect the movement.

``The wearing experience is comfortable. It is moderate and not squeezed. I was able to move normally during the wearing. '' ($\mathbf{P12}$)

``There is no inconvenience when wearing. The material is not thick, which is %
also %
acceptable during summer.'' ($\mathbf{P9}$)

Meanwhile, most of the users felt the tracking effect was %
accurate except  $\mathbf{P16}$, who pointed out there was %
slight shaking sometimes.

``Overall, the visualization effect is consistent with my movement. However, there is slight shaking in the visualization sometimes.''}

\subsection{Tracking Results with Different Numbers of Sensors}
\label{sec:num_sensors}
\begin{figure}[t]
    \centering
    \includegraphics[width=0.38\columnwidth]{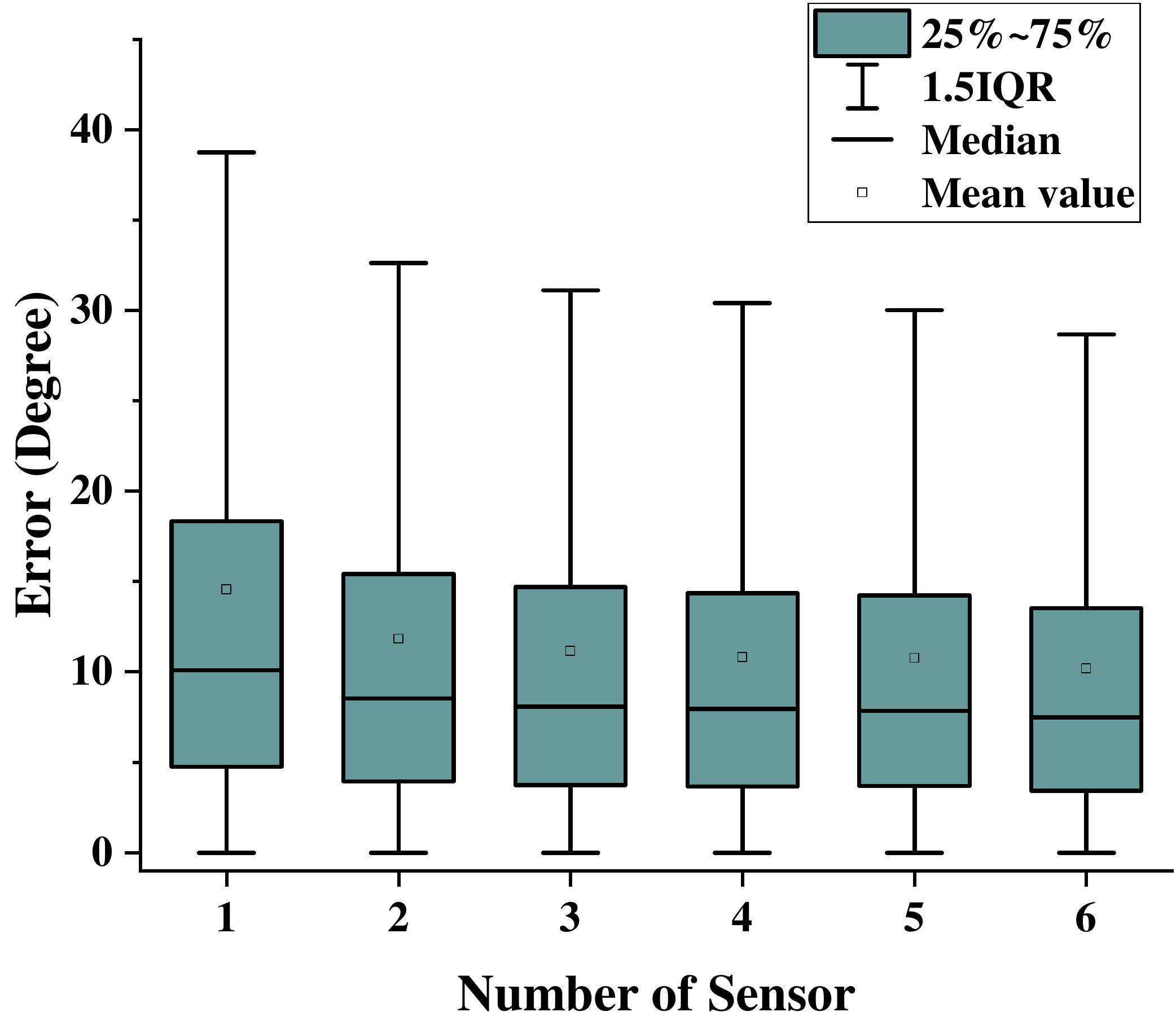}
      \caption{The average tracking errors change with the number of sensors.}~\label{fig:SensorNumber}
\end{figure}
\cg{The sensing capability of our prototype is rooted in the sensor stretching caused by the joint bending angle.
Fig.~\ref{fig:complexR} shows the sensor readings at different locations, indicating %
that the joint bending angle leads to a significant change in sensor readings of 2-3 sensors. 
These sensors are located at the semi-circle region on the side of the olecranon, and this region is effectively stretched when the joint rotates.
In contrast, the sensors on the side of the chelidon are not stretched and exhibit unnoticeable reading changes.
Our method should effectively cover the stretching semi-circle region given a full range of the circular displacement and up to 4 cm of the lateral displacement. Therefore, the number of sensors is expected to be 3-6.

We also numerically compare the results using different numbers of sensors (Fig.~\ref{fig:SensorNumber}). The reported error is an average over all the combinations of sensor choices. For example, when choosing five sensors for prediction, we randomly leave out one sensor, rebuild the prediction model, and repeat this process for six times. From these findings, we can conclude that the accuracy increases along with the number of sensors. However, the growth rate is lower when the number of sensors is four or five. This reveals that there is no significant change in accuracy at this level.  As a result, using three sensors is enough to predict joint angles with relatively high accuracy. }

\subsection{Comparison of Different Learning Methods}

We also compare our method with the sequential model and non-sequential model on the dataset $\mathcal{D}_{SS}$ without transfer learning. These methods include LSTM (with an extra batch normalization layer), LightGBM, \cg{The mix of CNN, LSTM (denoted as CL following)} and Fully-connected neural network (FCN). LSTM decreases the probability of gradient explosion and gradient disappearance. The mix of CNN and LSTM is a popular model handling human motion data \cite{singh2020deep}. LightGBM is determined as a powerful model which reduces the training time while keeping a good predicting ability \cite{ke2017lightgbm}. Fully-connected neural network (FCN) is a classical model in deep learning. Random forest \cite{breiman2001random} is an ensemble learning algorithm based on the decision tree. 
The comparison results are depicted in Table~\ref{tab:othermethod}.
They are both popular approaches in the domain of machine learning. 
It can be observed that the average tracking error and the variances of our method, CL, LightGBM, FCN and random forest are different from each other. While the LightGBM takes less training time, our method outperforms the other methods in average tracking error and the variance of the tracking errors while consuming only 98s to train. Since the other methods are not far ahead in training time, the mean tracking error does not exceed ours, we choose 
LSTM as our method.

\begin{table}[t]
      \caption{\cg{Comparisons of our method with alternative learning methods.}}~\label{tab:othermethod}
    \centering
    \begin{tabular}{cccccc}
       \toprule 
       Method  & Our Method & CL& LightGBM& FCN&Random Forest\\
       \midrule 
       Training time&  98s&422.6s& 1.2s &47s&378s\\
       \midrule 
       Run time& 0.001ms  &0.000001ms&0.019ms &0.001ms&0.316ms\\
       \midrule 
       Error on $\mathcal{D}_{SS-TE}$& 9.82 &11.40 & 15.91 &11.39&17.40\\
       \midrule
       Error on $\mathcal{D}_{SS-VA}$& 9.96 &12.17 & 16.04 &12.61&18.66\\
       \midrule 
       Variance on $\mathcal{D}_{SS-TE}$& 93.93& 161.50 & 137.31& 121.28&315.58\\
       \midrule 
       Variance on $\mathcal{D}_{SS-VA}$& 100.58& 184.71 & 158.19& 134.72&332.63\\
       \bottomrule 
    \end{tabular}

\end{table}

\subsection{Comparison of Different Ranking Methods}
\label{sec:rankingmethod}
\cg{To figure out the best strategy to re-arrange the data, we conduct an experiment with the ranking criterion of standard deviation, fuzzy entropy, jitter (indicating a lack of smoothness and naturalness, which is the third derivative of the position ~\cite{flash1985coordination}), and not ranking. We test them with the datasets of $\mathcal{D}_{SS}$, $\mathcal{D}_{MM}$ and $\mathcal{D}_{SM}$, respectively. 
\newcommand{\tabincell}[2]{\begin{tabular}{@{}#1@{}}#2\end{tabular}}
\begin{table}[t]
      \caption{Comparisons with alternative arrangement methods.}~\label{tab:otherarr}
    \centering
    \begin{tabular}{ccccc}
       \toprule 
	   & \tabincell{c}{no ranking\\(degree)}&\tabincell{c}{Jitter\\(10$^{4}$degree/s$^{3}$)}&\tabincell{c}{SD\\(degree)}& \tabincell{c}{Fuzzy Entropy\\(degree)}\\
 		\midrule
 		$\mathcal{D}_{SS-TE}$  & 9.82& 11.56&12.22& 10.66\\
 		 \midrule
 		$\mathcal{D}_{SS-VA}$  & 9.96& 12.48&13.86& 11.92\\
 		\midrule 
       $\mathcal{D}_{SM-TE}$&  18.82&14.51 &12.93&10.98\\
       \midrule 
      $\mathcal{D}_{SM-VA}$&  16.59&15.84 &11.25&10.27\\
       \midrule 
       $\mathcal{D}_{MM-TE}$&  27.96& 28.48 &18.95&11.81\\
       \midrule 
       $\mathcal{D}_{MM-VA}$&  29.34& 25.17 &16.06&12.53\\
       \midrule 
       Using time & 0s& 6.1*e-5s &5.5*e-5s & 0.022s\\
       \bottomrule

    \end{tabular}
\end{table}
From Table~\ref{tab:otherarr}, we can infer that fuzzy entropy outperforms the other %
methods on all the %
datasets except the $\mathcal{D}_{SS-TE}$ while \rev{the computing of fuzzy entropy} 
consumes the most time. That means when one user has collected his/her own dataset, there is no need to adopt the fuzzy entropy to displace the features. However, when the user wants to do other movements or there is a new user, it is necessary to adopt fuzzy entropy to adjust the features. }

\subsection{Latency Performance}
\cg{The total delay is about 40 ms, which can be divided into four parts. 
The first part is the sampling interval of the circuit board, which transmits data every 20 ms.
The second part is the process of the forward pass of the LSTM model estimating the joint angle, which causes a delay of 0.001 ms. 
The third part is the Kalman filter, which costs \rev{0.05} ms.%
The fourth part is the network delay. Since the server and the mobile Bluetooth transmitter are connected via a TCP network, the delay depends on the quality of the network signal. In our experiment (conducted using a local wireless area network), the delays at the server-mobile and server-computer stages are 2 ms and 17 ms, respectively. }

\subsection{Summary \& Discussions}

\cg{In conclusion, our method performs robustly over $\mathcal{D}_{SS}, \mathcal{D}_{SM}$ and $\mathcal{D}_{MM}$ with low latency. Besides, the sensors can also be reduced to three with low accuracy loss. Although the current elbow pad is limited to users whose arm \rev{girths are} between 20.5 cm and 28cm, it can be resolved by developing elbow pads with different sizes. In addition, the wearing experience is good across different users and it will not affect user movement whose arm \rev{girths are} between 20.5cm and 28cm. During the user evaluation interview, all the users also acknowledged the comfort of the DisPad. \rev{Besides, they thought the DisPad was accurate because the DisPad kept up with their movements throughout the experience.} However, as for the problem of jitters, there may be two possible reasons: (1) The elbow pad will produce some wrinkles during the data collecting process and real-time tracking, which induces the noise; (2) The model only fits the approximate functional relationship between sensor values and ground truth. As a result, there will be tracking errors during real-time use. These two problems are interesting, and solving them is meaningful for improving the effect of real-time tracking tools in future work. }

\section{Discussion}
\subsection{Limitations of Our Work}

In this work, we propose a learning-based method to tackle sensor displacement. Then, we validate it to multi-motions and multi-users. However, we found there is something that should be improved.

First, just detecting one DOF (degree of freedom) angle is not enough. While we assume that the elbow joint is a hinge constraint, i.e., it can only flex and extend around one axis; the actual elbow joint is far more complex than a single degree of freedom (DOF).
The observation is that the \emph{radius} rotates around the \emph{ulna} (Fig.~\ref{fig:arm}), allowing for forearm rotation.
Thus, to resolve this issue, detecting multiple joints is helpful to get sufficient information about the relevant joints, which is essential to analyze the complex signal patterns caused by these previously neglected DOFs. %
Second, when an existing user wants to perform a new type of motion, or there is a new user, we need to collect extra sensor data to do the transfer learning. However, compared to IMUs, which need calibration for every use, our method requires the collection of such data only once.

\subsection{Comparison with Other Fabric-Based Tracking Methods}
We discuss the difference between our technique and several state-of-the-art fabric-based tracking systems \cite{kim2018deep,esfahani2018smart,liu2019reconstructing}.
One study used a long short-term memory deep neural network to relate the sensor signal to full-body posture \cite{kim2018deep}.
They reported a tracking error between 1.2 to 5.7 centimeters in terms of the Euclidean distance between the estimated and baseline joint positions.
Their analysis was limited to three types of motion: squat, bend \& reach, and windmill. 
Another study investigated motion tracking for the torso and shoulder and achieved an average error of 9.4 degrees \cite{esfahani2018smart}.
While these two works achieve impressive tracking performance, they lack sufficient analysis of the sensitivity of the error to the factor of sensor displacement.
The recent study \cite{liu2019reconstructing}, which is the most relevant to our work, achieved an average tracking error of 9.69 degrees on the elbow joint.
They further investigated tracking performance when the sensor deviated from the ideal location by 1 cm, which caused the average tracking errors to increase by around 10 degrees (i.e., the errors went up to around 20 degrees.)
In comparison, we achieved 10.98 degrees across different motion types and 11.81 degrees across different users with transfer learning, given various configurations of sensor placements whose displacement is well beyond 1 cm.
This confirms the robustness of our method in tackling the issue of sensor displacement.

\section{Conclusion}
We have presented DisPad, which uses a sparse network of soft sensors on a textile pad to robustly estimate the elbow joint angle regardless of the variation in on-body device placement, different motions and different users. To deal with the sensor displacement, we leveraged an LSTM model to estimate the elbow joint angle. Besides, transfer learning is adopted to handle the different motion types and different users. \cg{To reduce the number of variables during transfer learning, we adopted the ranking based on fuzzy entropy.}
Then, we conducted comprehensive experiments to evaluate tracking errors across different users, motion states, and motion types and achieved an average tracking error of 9.82 degrees on the single-user fixed-motion dataset.
We further achieved stable tracking errors of 10.98 degrees and 11.81 degrees across different motion types and users, respectively, with the diverse on-body placement of our device prototype.

This work opens up a few directions for our future efforts.
The first one is to alleviate the demand for manual data collection.
We might resort to recent deep learning techniques (such as generative adversarial networks \cite{goodfellow2014generative}), which can be used to generate data for training.
The second direction is to extend our method to estimate full-body posture rather than just one elbow joint. 
It is common for athletes to wear both elbow and knee pads for protection, creating an ideal application scenario for our system.
Using four joint angles (two elbows and two knees) will be potentially sufficient to predict the angle of limb joints. 
The third research direction is to develop an application for patients who have impaired motor functions. 
Patients who have undergone fractures, paralytic strokes, or suffer from cerebral palsy face a lengthy period of rehabilitation, during which in-house self-training is key to their recovery progress.
Our method only requires wearing the device at an approximate location, and no expert knowledge is necessary, thus providing high convenience for users in need.

\bibliographystyle{ACM-Reference-Format}
\bibliography{sample-base}


\begin{thebibliography}{57}


\ifx \showCODEN    \undefined \def \showCODEN     #1{\unskip}     \fi
\ifx \showDOI      \undefined \def \showDOI       #1{#1}\fi
\ifx \showISBNx    \undefined \def \showISBNx     #1{\unskip}     \fi
\ifx \showISBNxiii \undefined \def \showISBNxiii  #1{\unskip}     \fi
\ifx \showISSN     \undefined \def \showISSN      #1{\unskip}     \fi
\ifx \showLCCN     \undefined \def \showLCCN      #1{\unskip}     \fi
\ifx \shownote     \undefined \def \shownote      #1{#1}          \fi
\ifx \showarticletitle \undefined \def \showarticletitle #1{#1}   \fi
\ifx \showURL      \undefined \def \showURL       {\relax}        \fi
\providecommand\bibfield[2]{#2}
\providecommand\bibinfo[2]{#2}
\providecommand\natexlab[1]{#1}
\providecommand\showeprint[2][]{arXiv:#2}

\bibitem[Ahuja et~al\mbox{.}(2021)]%
        {ahuja2021pose}
\bibfield{author}{\bibinfo{person}{Karan Ahuja}, \bibinfo{person}{Sven Mayer},
  \bibinfo{person}{Mayank Goel}, {and} \bibinfo{person}{Chris Harrison}.}
  \bibinfo{year}{2021}\natexlab{}.
\newblock \showarticletitle{Pose-on-the-Go: Approximating User Pose with
  Smartphone Sensor Fusion and Inverse Kinematics}. In
  \bibinfo{booktitle}{\emph{Proceedings of the 2021 CHI Conference on Human
  Factors in Computing Systems}}. \bibinfo{pages}{1--12}.
\newblock


\bibitem[Amjadi et~al\mbox{.}(2016)]%
        {amjadi2016stretchable}
\bibfield{author}{\bibinfo{person}{Morteza Amjadi}, \bibinfo{person}{Ki-Uk
  Kyung}, \bibinfo{person}{Inkyu Park}, {and} \bibinfo{person}{Metin Sitti}.}
  \bibinfo{year}{2016}\natexlab{}.
\newblock \showarticletitle{Stretchable, skin-mountable, and wearable strain
  sensors and their potential applications: a review}.
\newblock \bibinfo{journal}{\emph{Advanced Functional Materials}}
  \bibinfo{volume}{26}, \bibinfo{number}{11} (\bibinfo{year}{2016}),
  \bibinfo{pages}{1678--1698}.
\newblock


\bibitem[Bae et~al\mbox{.}(2013)]%
        {bae2013graphene}
\bibfield{author}{\bibinfo{person}{Sang-Hoon Bae}, \bibinfo{person}{Youngbin
  Lee}, \bibinfo{person}{Bhupendra~K Sharma}, \bibinfo{person}{Hak-Joo Lee},
  \bibinfo{person}{Jae-Hyun Kim}, {and} \bibinfo{person}{Jong-Hyun Ahn}.}
  \bibinfo{year}{2013}\natexlab{}.
\newblock \showarticletitle{Graphene-based transparent strain sensor}.
\newblock \bibinfo{journal}{\emph{Carbon}}  \bibinfo{volume}{51}
  (\bibinfo{year}{2013}), \bibinfo{pages}{236--242}.
\newblock


\bibitem[Banos et~al\mbox{.}(2014)]%
        {banos2014dealing}
\bibfield{author}{\bibinfo{person}{Oresti Banos}, \bibinfo{person}{Mate Toth},
  \bibinfo{person}{Miguel Damas}, \bibinfo{person}{Hector Pomares}, {and}
  \bibinfo{person}{Ignacio Rojas}.} \bibinfo{year}{2014}\natexlab{}.
\newblock \showarticletitle{Dealing with the effects of sensor displacement in
  wearable activity recognition}.
\newblock \bibinfo{journal}{\emph{Sensors}} \bibinfo{volume}{14},
  \bibinfo{number}{6} (\bibinfo{year}{2014}), \bibinfo{pages}{9995--10023}.
\newblock


\bibitem[Breiman(2001)]%
        {breiman2001random}
\bibfield{author}{\bibinfo{person}{Leo Breiman}.}
  \bibinfo{year}{2001}\natexlab{}.
\newblock \showarticletitle{Random forests}.
\newblock \bibinfo{journal}{\emph{Machine learning}} \bibinfo{volume}{45},
  \bibinfo{number}{1} (\bibinfo{year}{2001}), \bibinfo{pages}{5--32}.
\newblock


\bibitem[Chavarriaga et~al\mbox{.}(2013)]%
        {chavarriaga2013unsupervised}
\bibfield{author}{\bibinfo{person}{Ricardo Chavarriaga},
  \bibinfo{person}{Hamidreza Bayati}, {and} \bibinfo{person}{Jos{\'e}~Del
  Mill{\'a}n}.} \bibinfo{year}{2013}\natexlab{}.
\newblock \showarticletitle{Unsupervised adaptation for acceleration-based
  activity recognition: robustness to sensor displacement and rotation}.
\newblock \bibinfo{journal}{\emph{Personal and Ubiquitous Computing}}
  \bibinfo{volume}{17}, \bibinfo{number}{3} (\bibinfo{year}{2013}),
  \bibinfo{pages}{479--490}.
\newblock


\bibitem[Chen et~al\mbox{.}(2012)]%
        {chen2012sensor}
\bibfield{author}{\bibinfo{person}{Liming Chen}, \bibinfo{person}{Jesse Hoey},
  \bibinfo{person}{Chris~D Nugent}, \bibinfo{person}{Diane~J Cook}, {and}
  \bibinfo{person}{Zhiwen Yu}.} \bibinfo{year}{2012}\natexlab{}.
\newblock \showarticletitle{Sensor-based activity recognition}.
\newblock \bibinfo{journal}{\emph{IEEE Transactions on Systems, Man, and
  Cybernetics, Part C (Applications and Reviews)}} \bibinfo{volume}{42},
  \bibinfo{number}{6} (\bibinfo{year}{2012}), \bibinfo{pages}{790--808}.
\newblock


\bibitem[Chen et~al\mbox{.}(2007a)]%
        {chen2007real}
\bibfield{author}{\bibinfo{person}{Qing Chen}, \bibinfo{person}{Nicolas~D
  Georganas}, {and} \bibinfo{person}{Emil~M Petriu}.}
  \bibinfo{year}{2007}\natexlab{a}.
\newblock \showarticletitle{Real-time vision-based hand gesture recognition
  using haar-like features}. In \bibinfo{booktitle}{\emph{2007 IEEE
  instrumentation \& measurement technology conference IMTC 2007}}. IEEE,
  \bibinfo{pages}{1--6}.
\newblock


\bibitem[Chen et~al\mbox{.}(2007b)]%
        {chen2007characterization}
\bibfield{author}{\bibinfo{person}{Weiting Chen}, \bibinfo{person}{Zhizhong
  Wang}, \bibinfo{person}{Hongbo Xie}, {and} \bibinfo{person}{Wangxin Yu}.}
  \bibinfo{year}{2007}\natexlab{b}.
\newblock \showarticletitle{Characterization of surface EMG signal based on
  fuzzy entropy}.
\newblock \bibinfo{journal}{\emph{IEEE Transactions on neural systems and
  rehabilitation engineering}} \bibinfo{volume}{15}, \bibinfo{number}{2}
  (\bibinfo{year}{2007}), \bibinfo{pages}{266--272}.
\newblock


\bibitem[Chossat et~al\mbox{.}(2015)]%
        {chossat2015wearable}
\bibfield{author}{\bibinfo{person}{Jean-Baptiste Chossat},
  \bibinfo{person}{Yiwei Tao}, \bibinfo{person}{Vincent Duchaine}, {and}
  \bibinfo{person}{Yong-Lae Park}.} \bibinfo{year}{2015}\natexlab{}.
\newblock \showarticletitle{Wearable soft artificial skin for hand motion
  detection with embedded microfluidic strain sensing}. In
  \bibinfo{booktitle}{\emph{2015 IEEE international conference on robotics and
  automation (ICRA)}}. IEEE, \bibinfo{pages}{2568--2573}.
\newblock


\bibitem[Cornacchia et~al\mbox{.}(2016)]%
        {cornacchia2016survey}
\bibfield{author}{\bibinfo{person}{Maria Cornacchia}, \bibinfo{person}{Koray
  Ozcan}, \bibinfo{person}{Yu Zheng}, {and} \bibinfo{person}{Senem
  Velipasalar}.} \bibinfo{year}{2016}\natexlab{}.
\newblock \showarticletitle{A survey on activity detection and classification
  using wearable sensors}.
\newblock \bibinfo{journal}{\emph{IEEE Sensors Journal}} \bibinfo{volume}{17},
  \bibinfo{number}{2} (\bibinfo{year}{2016}), \bibinfo{pages}{386--403}.
\newblock


\bibitem[Einsmann et~al\mbox{.}(2005)]%
        {einsmann2005modeling}
\bibfield{author}{\bibinfo{person}{Christopher Einsmann},
  \bibinfo{person}{Meghan Quirk}, \bibinfo{person}{Ben Muzal},
  \bibinfo{person}{Bharath Venkatramani}, \bibinfo{person}{Thomas Martin},
  {and} \bibinfo{person}{Mark Jones}.} \bibinfo{year}{2005}\natexlab{}.
\newblock \showarticletitle{Modeling a wearable full-body motion capture
  system}. In \bibinfo{booktitle}{\emph{Ninth IEEE International Symposium on
  Wearable Computers (ISWC'05)}}. IEEE, \bibinfo{pages}{144--151}.
\newblock


\bibitem[Enokibori and Mase(2014)]%
        {enokibori2014human}
\bibfield{author}{\bibinfo{person}{Yu Enokibori} {and} \bibinfo{person}{Kenji
  Mase}.} \bibinfo{year}{2014}\natexlab{}.
\newblock \showarticletitle{Human joint angle estimation with an e-textile
  sensor}. In \bibinfo{booktitle}{\emph{Proceedings of the 2014 ACM
  International Symposium on Wearable Computers}}. \bibinfo{pages}{129--130}.
\newblock


\bibitem[Esfahani and Nussbaum(2018)]%
        {esfahani2018smart}
\bibfield{author}{\bibinfo{person}{Mohammad Iman~Mokhlespour Esfahani} {and}
  \bibinfo{person}{Maury~A Nussbaum}.} \bibinfo{year}{2018}\natexlab{}.
\newblock \showarticletitle{A “smart” undershirt for tracking upper body
  motions: Task classification and angle estimation}.
\newblock \bibinfo{journal}{\emph{IEEE sensors Journal}} \bibinfo{volume}{18},
  \bibinfo{number}{18} (\bibinfo{year}{2018}), \bibinfo{pages}{7650--7658}.
\newblock


\bibitem[Farringdon et~al\mbox{.}(1999)]%
        {farringdon1999wearable}
\bibfield{author}{\bibinfo{person}{Jonny Farringdon}, \bibinfo{person}{Andrew~J
  Moore}, \bibinfo{person}{Nancy Tilbury}, \bibinfo{person}{James Church},
  {and} \bibinfo{person}{Pieter~D Biemond}.} \bibinfo{year}{1999}\natexlab{}.
\newblock \showarticletitle{Wearable sensor badge and sensor jacket for context
  awareness}. In \bibinfo{booktitle}{\emph{Digest of Papers. Third
  International Symposium on Wearable Computers}}. IEEE,
  \bibinfo{pages}{107--113}.
\newblock


\bibitem[Filippeschi et~al\mbox{.}(2017)]%
        {filippeschi2017survey}
\bibfield{author}{\bibinfo{person}{Alessandro Filippeschi},
  \bibinfo{person}{Norbert Schmitz}, \bibinfo{person}{Markus Miezal},
  \bibinfo{person}{Gabriele Bleser}, \bibinfo{person}{Emanuele Ruffaldi}, {and}
  \bibinfo{person}{Didier Stricker}.} \bibinfo{year}{2017}\natexlab{}.
\newblock \showarticletitle{Survey of motion tracking methods based on inertial
  sensors: A focus on upper limb human motion}.
\newblock \bibinfo{journal}{\emph{Sensors}} \bibinfo{volume}{17},
  \bibinfo{number}{6} (\bibinfo{year}{2017}), \bibinfo{pages}{1257}.
\newblock


\bibitem[Flash and Hogan(1985)]%
        {flash1985coordination}
\bibfield{author}{\bibinfo{person}{Tamar Flash} {and} \bibinfo{person}{Neville
  Hogan}.} \bibinfo{year}{1985}\natexlab{}.
\newblock \showarticletitle{The coordination of arm movements: an
  experimentally confirmed mathematical model}.
\newblock \bibinfo{journal}{\emph{Journal of neuroscience}}
  \bibinfo{volume}{5}, \bibinfo{number}{7} (\bibinfo{year}{1985}),
  \bibinfo{pages}{1688--1703}.
\newblock


\bibitem[F{\"o}rster et~al\mbox{.}(2009)]%
        {forster2009evolving}
\bibfield{author}{\bibinfo{person}{Kilian F{\"o}rster}, \bibinfo{person}{Pascal
  Brem}, \bibinfo{person}{Daniel Roggen}, {and} \bibinfo{person}{Gerhard
  Tr{\"o}ster}.} \bibinfo{year}{2009}\natexlab{}.
\newblock \showarticletitle{Evolving discriminative features robust to sensor
  displacement for activity recognition in body area sensor networks}. In
  \bibinfo{booktitle}{\emph{2009 International Conference on Intelligent
  Sensors, Sensor Networks and Information Processing (ISSNIP)}}. IEEE,
  \bibinfo{pages}{43--48}.
\newblock


\bibitem[Forster et~al\mbox{.}(2009)]%
        {forster2009unsupervised}
\bibfield{author}{\bibinfo{person}{Kilian Forster}, \bibinfo{person}{Daniel
  Roggen}, {and} \bibinfo{person}{Gerhard Troster}.}
  \bibinfo{year}{2009}\natexlab{}.
\newblock \showarticletitle{Unsupervised classifier self-calibration through
  repeated context occurences: Is there robustness against sensor displacement
  to gain?}. In \bibinfo{booktitle}{\emph{2009 international symposium on
  wearable computers}}. IEEE, \bibinfo{pages}{77--84}.
\newblock


\bibitem[Gibbs and Asada(2005)]%
        {gibbs2005wearable}
\bibfield{author}{\bibinfo{person}{Peter~T Gibbs} {and} \bibinfo{person}{HHarry
  Asada}.} \bibinfo{year}{2005}\natexlab{}.
\newblock \showarticletitle{Wearable conductive fiber sensors for multi-axis
  human joint angle measurements}.
\newblock \bibinfo{journal}{\emph{Journal of neuroengineering and
  rehabilitation}} \bibinfo{volume}{2}, \bibinfo{number}{1}
  (\bibinfo{year}{2005}), \bibinfo{pages}{7}.
\newblock


\bibitem[Gioberto and Dunne(2012)]%
        {gioberto2012garment}
\bibfield{author}{\bibinfo{person}{Guido Gioberto} {and}
  \bibinfo{person}{Lucy~E Dunne}.} \bibinfo{year}{2012}\natexlab{}.
\newblock \showarticletitle{Garment positioning and drift in garment-integrated
  wearable sensing}. In \bibinfo{booktitle}{\emph{2012 16th International
  symposium on wearable computers}}. IEEE, \bibinfo{pages}{64--71}.
\newblock


\bibitem[Glauser et~al\mbox{.}(2019a)]%
        {glauser2019deformation}
\bibfield{author}{\bibinfo{person}{Oliver Glauser}, \bibinfo{person}{Daniele
  Panozzo}, \bibinfo{person}{Otmar Hilliges}, {and} \bibinfo{person}{Olga
  Sorkine-Hornung}.} \bibinfo{year}{2019}\natexlab{a}.
\newblock \showarticletitle{Deformation capture via soft and stretchable sensor
  arrays}.
\newblock \bibinfo{journal}{\emph{ACM Transactions on Graphics (TOG)}}
  \bibinfo{volume}{38}, \bibinfo{number}{2} (\bibinfo{year}{2019}),
  \bibinfo{pages}{16}.
\newblock


\bibitem[Glauser et~al\mbox{.}(2019b)]%
        {glauser2019interactive}
\bibfield{author}{\bibinfo{person}{Oliver Glauser}, \bibinfo{person}{Shihao
  Wu}, \bibinfo{person}{Daniele Panozzo}, \bibinfo{person}{Otmar Hilliges},
  {and} \bibinfo{person}{Olga Sorkine-Hornung}.}
  \bibinfo{year}{2019}\natexlab{b}.
\newblock \showarticletitle{Interactive hand pose estimation using a
  stretch-sensing soft glove}.
\newblock \bibinfo{journal}{\emph{ACM Transactions on Graphics (TOG)}}
  \bibinfo{volume}{38}, \bibinfo{number}{4} (\bibinfo{year}{2019}),
  \bibinfo{pages}{41}.
\newblock


\bibitem[Gong et~al\mbox{.}(2015)]%
        {gong2015highly}
\bibfield{author}{\bibinfo{person}{Shu Gong}, \bibinfo{person}{Daniel~TH Lai},
  \bibinfo{person}{Bin Su}, \bibinfo{person}{Kae~Jye Si},
  \bibinfo{person}{Zheng Ma}, \bibinfo{person}{Lim~Wei Yap},
  \bibinfo{person}{Pengzhen Guo}, {and} \bibinfo{person}{Wenlong Cheng}.}
  \bibinfo{year}{2015}\natexlab{}.
\newblock \showarticletitle{Highly Stretchy Black Gold E-Skin Nanopatches as
  Highly Sensitive Wearable Biomedical Sensors}.
\newblock \bibinfo{journal}{\emph{Advanced Electronic Materials}}
  \bibinfo{volume}{1}, \bibinfo{number}{4} (\bibinfo{year}{2015}),
  \bibinfo{pages}{1400063}.
\newblock


\bibitem[Goodfellow et~al\mbox{.}(2014)]%
        {goodfellow2014generative}
\bibfield{author}{\bibinfo{person}{Ian Goodfellow}, \bibinfo{person}{Jean
  Pouget-Abadie}, \bibinfo{person}{Mehdi Mirza}, \bibinfo{person}{Bing Xu},
  \bibinfo{person}{David Warde-Farley}, \bibinfo{person}{Sherjil Ozair},
  \bibinfo{person}{Aaron Courville}, {and} \bibinfo{person}{Yoshua Bengio}.}
  \bibinfo{year}{2014}\natexlab{}.
\newblock \showarticletitle{Generative adversarial nets}. In
  \bibinfo{booktitle}{\emph{Advances in neural information processing
  systems}}. \bibinfo{pages}{2672--2680}.
\newblock


\bibitem[Gretton et~al\mbox{.}(2012a)]%
        {gretton2012kernel}
\bibfield{author}{\bibinfo{person}{Arthur Gretton}, \bibinfo{person}{Karsten~M
  Borgwardt}, \bibinfo{person}{Malte~J Rasch}, \bibinfo{person}{Bernhard
  Sch{\"o}lkopf}, {and} \bibinfo{person}{Alexander Smola}.}
  \bibinfo{year}{2012}\natexlab{a}.
\newblock \showarticletitle{A kernel two-sample test}.
\newblock \bibinfo{journal}{\emph{The Journal of Machine Learning Research}}
  \bibinfo{volume}{13}, \bibinfo{number}{1} (\bibinfo{year}{2012}),
  \bibinfo{pages}{723--773}.
\newblock


\bibitem[Gretton et~al\mbox{.}(2012b)]%
        {gretton2012optimal}
\bibfield{author}{\bibinfo{person}{Arthur Gretton}, \bibinfo{person}{Dino
  Sejdinovic}, \bibinfo{person}{Heiko Strathmann}, \bibinfo{person}{Sivaraman
  Balakrishnan}, \bibinfo{person}{Massimiliano Pontil}, \bibinfo{person}{Kenji
  Fukumizu}, {and} \bibinfo{person}{Bharath~K Sriperumbudur}.}
  \bibinfo{year}{2012}\natexlab{b}.
\newblock \showarticletitle{Optimal kernel choice for large-scale two-sample
  tests}.
\newblock \bibinfo{journal}{\emph{Advances in neural information processing
  systems}}  \bibinfo{volume}{25} (\bibinfo{year}{2012}).
\newblock


\bibitem[Haratian et~al\mbox{.}(2013)]%
        {haratian2013toward}
\bibfield{author}{\bibinfo{person}{Roya Haratian}, \bibinfo{person}{Richard
  Twycross-Lewis}, \bibinfo{person}{Tijana Timotijevic}, {and}
  \bibinfo{person}{Chris Phillips}.} \bibinfo{year}{2013}\natexlab{}.
\newblock \showarticletitle{Toward flexibility in sensor placement for motion
  capture systems: a signal processing approach}.
\newblock \bibinfo{journal}{\emph{IEEE Sensors Journal}} \bibinfo{volume}{14},
  \bibinfo{number}{3} (\bibinfo{year}{2013}), \bibinfo{pages}{701--709}.
\newblock


\bibitem[Hochreiter and Schmidhuber(1997)]%
        {hochreiter1997long}
\bibfield{author}{\bibinfo{person}{Sepp Hochreiter} {and}
  \bibinfo{person}{J{\"u}rgen Schmidhuber}.} \bibinfo{year}{1997}\natexlab{}.
\newblock \showarticletitle{Long short-term memory}.
\newblock \bibinfo{journal}{\emph{Neural computation}} \bibinfo{volume}{9},
  \bibinfo{number}{8} (\bibinfo{year}{1997}), \bibinfo{pages}{1735--1780}.
\newblock


\bibitem[Huang et~al\mbox{.}(2018)]%
        {huang2018deep}
\bibfield{author}{\bibinfo{person}{Yinghao Huang}, \bibinfo{person}{Manuel
  Kaufmann}, \bibinfo{person}{Emre Aksan}, \bibinfo{person}{Michael~J Black},
  \bibinfo{person}{Otmar Hilliges}, {and} \bibinfo{person}{Gerard Pons-Moll}.}
  \bibinfo{year}{2018}\natexlab{}.
\newblock \showarticletitle{Deep inertial poser: learning to reconstruct human
  pose from sparse inertial measurements in real time}. In
  \bibinfo{booktitle}{\emph{SIGGRAPH Asia 2018 Technical Papers}}. ACM,
  \bibinfo{pages}{185}.
\newblock


\bibitem[Kale and Patil(2016)]%
        {kale2016study}
\bibfield{author}{\bibinfo{person}{Geetanjali~Vinayak Kale} {and}
  \bibinfo{person}{Varsha~Hemant Patil}.} \bibinfo{year}{2016}\natexlab{}.
\newblock \showarticletitle{A study of vision based human motion recognition
  and analysis}.
\newblock \bibinfo{journal}{\emph{International Journal of Ambient Computing
  and Intelligence (IJACI)}} \bibinfo{volume}{7}, \bibinfo{number}{2}
  (\bibinfo{year}{2016}), \bibinfo{pages}{75--92}.
\newblock


\bibitem[Ke et~al\mbox{.}(2017)]%
        {ke2017lightgbm}
\bibfield{author}{\bibinfo{person}{Guolin Ke}, \bibinfo{person}{Qi Meng},
  \bibinfo{person}{Thomas Finley}, \bibinfo{person}{Taifeng Wang},
  \bibinfo{person}{Wei Chen}, \bibinfo{person}{Weidong Ma},
  \bibinfo{person}{Qiwei Ye}, {and} \bibinfo{person}{Tie-Yan Liu}.}
  \bibinfo{year}{2017}\natexlab{}.
\newblock \showarticletitle{Lightgbm: A highly efficient gradient boosting
  decision tree}. In \bibinfo{booktitle}{\emph{Advances in neural information
  processing systems}}. \bibinfo{pages}{3146--3154}.
\newblock


\bibitem[Kiaghadi et~al\mbox{.}(2018)]%
        {kiaghadi2018fabric}
\bibfield{author}{\bibinfo{person}{Ali Kiaghadi}, \bibinfo{person}{Morgan
  Baima}, \bibinfo{person}{Jeremy Gummeson}, \bibinfo{person}{Trisha Andrew},
  {and} \bibinfo{person}{Deepak Ganesan}.} \bibinfo{year}{2018}\natexlab{}.
\newblock \showarticletitle{Fabric as a Sensor: Towards Unobtrusive Sensing of
  Human Behavior with Triboelectric Textiles}. In
  \bibinfo{booktitle}{\emph{Proceedings of the 16th ACM Conference on Embedded
  Networked Sensor Systems}}. ACM, \bibinfo{pages}{199--210}.
\newblock


\bibitem[Kim et~al\mbox{.}(2018)]%
        {kim2018deep}
\bibfield{author}{\bibinfo{person}{Dooyoung Kim}, \bibinfo{person}{Junghan
  Kwon}, \bibinfo{person}{Seunghyun Han}, \bibinfo{person}{Yong-Lae Park},
  {and} \bibinfo{person}{Sungho Jo}.} \bibinfo{year}{2018}\natexlab{}.
\newblock \showarticletitle{Deep full-body motion network for a soft wearable
  motion sensing suit}.
\newblock \bibinfo{journal}{\emph{IEEE/ASME Transactions on Mechatronics}}
  \bibinfo{volume}{24}, \bibinfo{number}{1} (\bibinfo{year}{2018}),
  \bibinfo{pages}{56--66}.
\newblock


\bibitem[Kunze and Lukowicz(2008)]%
        {kunze2008dealing}
\bibfield{author}{\bibinfo{person}{Kai Kunze} {and} \bibinfo{person}{Paul
  Lukowicz}.} \bibinfo{year}{2008}\natexlab{}.
\newblock \showarticletitle{Dealing with sensor displacement in motion-based
  onbody activity recognition systems}. In
  \bibinfo{booktitle}{\emph{Proceedings of the 10th international conference on
  Ubiquitous computing}}. ACM, \bibinfo{pages}{20--29}.
\newblock


\bibitem[Kunze and Lukowicz(2014)]%
        {kunze2014sensor}
\bibfield{author}{\bibinfo{person}{Kai Kunze} {and} \bibinfo{person}{Paul
  Lukowicz}.} \bibinfo{year}{2014}\natexlab{}.
\newblock \showarticletitle{Sensor placement variations in wearable activity
  recognition}.
\newblock \bibinfo{journal}{\emph{IEEE Pervasive Computing}}
  \bibinfo{volume}{13}, \bibinfo{number}{4} (\bibinfo{year}{2014}),
  \bibinfo{pages}{32--41}.
\newblock


\bibitem[Lee et~al\mbox{.}(2016)]%
        {lee2016novel}
\bibfield{author}{\bibinfo{person}{Sunghoon~Ivan Lee},
  \bibinfo{person}{Jean-Francois Daneault}, \bibinfo{person}{Luc Weydert},
  {and} \bibinfo{person}{Paolo Bonato}.} \bibinfo{year}{2016}\natexlab{}.
\newblock \showarticletitle{A novel flexible wearable sensor for estimating
  joint-angles}. In \bibinfo{booktitle}{\emph{2016 IEEE 13th International
  Conference on Wearable and Implantable Body Sensor Networks (BSN)}}. IEEE,
  \bibinfo{pages}{377--382}.
\newblock


\bibitem[Liu et~al\mbox{.}(2019)]%
        {liu2019reconstructing}
\bibfield{author}{\bibinfo{person}{Ruibo Liu}, \bibinfo{person}{Qijia Shao},
  \bibinfo{person}{Siqi Wang}, \bibinfo{person}{Christina Ru},
  \bibinfo{person}{Devin Balkcom}, {and} \bibinfo{person}{Xia Zhou}.}
  \bibinfo{year}{2019}\natexlab{}.
\newblock \showarticletitle{Reconstructing Human Joint Motion with
  Computational Fabrics}.
\newblock \bibinfo{journal}{\emph{Proceedings of the ACM on Interactive,
  Mobile, Wearable and Ubiquitous Technologies}} \bibinfo{volume}{3},
  \bibinfo{number}{1} (\bibinfo{year}{2019}), \bibinfo{pages}{19}.
\newblock


\bibitem[Lorussi et~al\mbox{.}(2005)]%
        {lorussi2005strain}
\bibfield{author}{\bibinfo{person}{Federico Lorussi},
  \bibinfo{person}{Enzo~Pasquale Scilingo}, \bibinfo{person}{Mario Tesconi},
  \bibinfo{person}{Alessandro Tognetti}, {and} \bibinfo{person}{Danilo
  De~Rossi}.} \bibinfo{year}{2005}\natexlab{}.
\newblock \showarticletitle{Strain sensing fabric for hand posture and gesture
  monitoring}.
\newblock \bibinfo{journal}{\emph{IEEE Transactions on Information Technology
  in Biomedicine}} \bibinfo{volume}{9}, \bibinfo{number}{3}
  (\bibinfo{year}{2005}), \bibinfo{pages}{372--381}.
\newblock


\bibitem[Meng{\"u}{\c{c}} et~al\mbox{.}(2014)]%
        {mengucc2014wearable}
\bibfield{author}{\bibinfo{person}{Yi{\u{g}}it Meng{\"u}{\c{c}}},
  \bibinfo{person}{Yong-Lae Park}, \bibinfo{person}{Hao Pei},
  \bibinfo{person}{Daniel Vogt}, \bibinfo{person}{Patrick~M Aubin},
  \bibinfo{person}{Ethan Winchell}, \bibinfo{person}{Lowell Fluke},
  \bibinfo{person}{Leia Stirling}, \bibinfo{person}{Robert~J Wood}, {and}
  \bibinfo{person}{Conor~J Walsh}.} \bibinfo{year}{2014}\natexlab{}.
\newblock \showarticletitle{Wearable soft sensing suit for human gait
  measurement}.
\newblock \bibinfo{journal}{\emph{The International Journal of Robotics
  Research}} \bibinfo{volume}{33}, \bibinfo{number}{14} (\bibinfo{year}{2014}),
  \bibinfo{pages}{1748--1764}.
\newblock


\bibitem[Meyer et~al\mbox{.}(2006)]%
        {meyer2006textile}
\bibfield{author}{\bibinfo{person}{Jan Meyer}, \bibinfo{person}{Paul Lukowicz},
  {and} \bibinfo{person}{Gerhard Troster}.} \bibinfo{year}{2006}\natexlab{}.
\newblock \showarticletitle{Textile pressure sensor for muscle activity and
  motion detection}. In \bibinfo{booktitle}{\emph{2006 10th IEEE International
  Symposium on Wearable Computers}}. IEEE, \bibinfo{pages}{69--72}.
\newblock


\bibitem[Poupyrev et~al\mbox{.}(2016)]%
        {poupyrev2016project}
\bibfield{author}{\bibinfo{person}{Ivan Poupyrev}, \bibinfo{person}{Nan-Wei
  Gong}, \bibinfo{person}{Shiho Fukuhara}, \bibinfo{person}{Mustafa~Emre
  Karagozler}, \bibinfo{person}{Carsten Schwesig}, {and}
  \bibinfo{person}{Karen~E Robinson}.} \bibinfo{year}{2016}\natexlab{}.
\newblock \showarticletitle{Project Jacquard: interactive digital textiles at
  scale}. In \bibinfo{booktitle}{\emph{Proceedings of the 2016 CHI Conference
  on Human Factors in Computing Systems}}. \bibinfo{pages}{4216--4227}.
\newblock


\bibitem[Santoyo(2017)]%
        {santoyo2017brief}
\bibfield{author}{\bibinfo{person}{Sergio Santoyo}.}
  \bibinfo{year}{2017}\natexlab{}.
\newblock \showarticletitle{A brief overview of outlier detection techniques}.
\newblock \bibinfo{journal}{\emph{Towards data sciemce}}
  (\bibinfo{year}{2017}).
\newblock


\bibitem[Schneegass et~al\mbox{.}(2014)]%
        {schneegass2014towards}
\bibfield{author}{\bibinfo{person}{Stefan Schneegass}, \bibinfo{person}{Mariam
  Hassib}, \bibinfo{person}{Tobias Birmili}, {and} \bibinfo{person}{Niels
  Henze}.} \bibinfo{year}{2014}\natexlab{}.
\newblock \showarticletitle{Towards a garment OS: supporting application
  development for smart garments}. In \bibinfo{booktitle}{\emph{Proceedings of
  the 2014 ACM international symposium on wearable computers: Adjunct
  program}}. \bibinfo{pages}{261--266}.
\newblock


\bibitem[Schneegass et~al\mbox{.}(2015)]%
        {schneegass2015simpleskin}
\bibfield{author}{\bibinfo{person}{Stefan Schneegass}, \bibinfo{person}{Mariam
  Hassib}, \bibinfo{person}{Bo Zhou}, \bibinfo{person}{Jingyuan Cheng},
  \bibinfo{person}{Fernando Seoane}, \bibinfo{person}{Oliver Amft},
  \bibinfo{person}{Paul Lukowicz}, {and} \bibinfo{person}{Albrecht Schmidt}.}
  \bibinfo{year}{2015}\natexlab{}.
\newblock \showarticletitle{SimpleSkin: towards multipurpose smart garments}.
  In \bibinfo{booktitle}{\emph{Adjunct Proceedings of the 2015 ACM
  International Joint Conference on Pervasive and Ubiquitous Computing and
  Proceedings of the 2015 ACM International Symposium on Wearable Computers}}.
  \bibinfo{pages}{241--244}.
\newblock


\bibitem[Singh et~al\mbox{.}(2020)]%
        {singh2020deep}
\bibfield{author}{\bibinfo{person}{Satya~P Singh}, \bibinfo{person}{Madan~Kumar
  Sharma}, \bibinfo{person}{Aim{\'e} Lay-Ekuakille}, \bibinfo{person}{Deepak
  Gangwar}, {and} \bibinfo{person}{Sukrit Gupta}.}
  \bibinfo{year}{2020}\natexlab{}.
\newblock \showarticletitle{Deep ConvLSTM with self-attention for human
  activity decoding using wearable sensors}.
\newblock \bibinfo{journal}{\emph{IEEE Sensors Journal}} \bibinfo{volume}{21},
  \bibinfo{number}{6} (\bibinfo{year}{2020}), \bibinfo{pages}{8575--8582}.
\newblock


\bibitem[Someya and Amagai(2019)]%
        {someya2019toward}
\bibfield{author}{\bibinfo{person}{Takao Someya} {and}
  \bibinfo{person}{Masayuki Amagai}.} \bibinfo{year}{2019}\natexlab{}.
\newblock \showarticletitle{Toward a new generation of smart skins}.
\newblock \bibinfo{journal}{\emph{Nature biotechnology}} \bibinfo{volume}{37},
  \bibinfo{number}{4} (\bibinfo{year}{2019}), \bibinfo{pages}{382}.
\newblock


\bibitem[Steven~Eyobu and Han(2018)]%
        {steven2018feature}
\bibfield{author}{\bibinfo{person}{Odongo Steven~Eyobu} {and}
  \bibinfo{person}{Dong~Seog Han}.} \bibinfo{year}{2018}\natexlab{}.
\newblock \showarticletitle{Feature representation and data augmentation for
  human activity classification based on wearable IMU sensor data using a deep
  LSTM neural network}.
\newblock \bibinfo{journal}{\emph{Sensors}} \bibinfo{volume}{18},
  \bibinfo{number}{9} (\bibinfo{year}{2018}), \bibinfo{pages}{2892}.
\newblock


\bibitem[Sugiyama et~al\mbox{.}(2007)]%
        {sugiyama2007covariate}
\bibfield{author}{\bibinfo{person}{Masashi Sugiyama}, \bibinfo{person}{Matthias
  Krauledat}, {and} \bibinfo{person}{Klaus-Robert M{\~A}{\v{z}}ller}.}
  \bibinfo{year}{2007}\natexlab{}.
\newblock \showarticletitle{Covariate shift adaptation by importance weighted
  cross validation}.
\newblock \bibinfo{journal}{\emph{Journal of Machine Learning Research}}
  \bibinfo{volume}{8}, \bibinfo{number}{May} (\bibinfo{year}{2007}),
  \bibinfo{pages}{985--1005}.
\newblock


\bibitem[Tian et~al\mbox{.}(2015)]%
        {tian2015upper}
\bibfield{author}{\bibinfo{person}{Yushuang Tian}, \bibinfo{person}{Xiaoli
  Meng}, \bibinfo{person}{Dapeng Tao}, \bibinfo{person}{Dongquan Liu}, {and}
  \bibinfo{person}{Chen Feng}.} \bibinfo{year}{2015}\natexlab{}.
\newblock \showarticletitle{Upper limb motion tracking with the integration of
  IMU and Kinect}.
\newblock \bibinfo{journal}{\emph{Neurocomputing}}  \bibinfo{volume}{159}
  (\bibinfo{year}{2015}), \bibinfo{pages}{207--218}.
\newblock


\bibitem[Tognetti et~al\mbox{.}(2014)]%
        {tognetti2014new}
\bibfield{author}{\bibinfo{person}{Alessandro Tognetti},
  \bibinfo{person}{Federico Lorussi}, \bibinfo{person}{Gabriele Dalle~Mura},
  \bibinfo{person}{Nicola Carbonaro}, \bibinfo{person}{Maria Pacelli},
  \bibinfo{person}{Rita Paradiso}, {and} \bibinfo{person}{Danilo De~Rossi}.}
  \bibinfo{year}{2014}\natexlab{}.
\newblock \showarticletitle{New generation of wearable goniometers for motion
  capture systems}.
\newblock \bibinfo{journal}{\emph{Journal of neuroengineering and
  rehabilitation}} \bibinfo{volume}{11}, \bibinfo{number}{1}
  (\bibinfo{year}{2014}), \bibinfo{pages}{56}.
\newblock


\bibitem[Voit and Schneegass(2017)]%
        {voit2017fabricid}
\bibfield{author}{\bibinfo{person}{Alexandra Voit} {and}
  \bibinfo{person}{Stefan Schneegass}.} \bibinfo{year}{2017}\natexlab{}.
\newblock \showarticletitle{FabricID: Using smart textiles to access wearable
  devices}. In \bibinfo{booktitle}{\emph{Proceedings of the 16th International
  Conference on Mobile and Ubiquitous Multimedia}}. \bibinfo{pages}{379--385}.
\newblock


\bibitem[von Marcard et~al\mbox{.}(2018)]%
        {von2018recovering}
\bibfield{author}{\bibinfo{person}{Timo von Marcard}, \bibinfo{person}{Roberto
  Henschel}, \bibinfo{person}{Michael~J Black}, \bibinfo{person}{Bodo
  Rosenhahn}, {and} \bibinfo{person}{Gerard Pons-Moll}.}
  \bibinfo{year}{2018}\natexlab{}.
\newblock \showarticletitle{Recovering accurate 3d human pose in the wild using
  imus and a moving camera}. In \bibinfo{booktitle}{\emph{Proceedings of the
  European Conference on Computer Vision (ECCV)}}. \bibinfo{pages}{601--617}.
\newblock


\bibitem[Xiao and Zarar(2018)]%
        {xiao2018machine}
\bibfield{author}{\bibinfo{person}{Xuesu Xiao} {and} \bibinfo{person}{Shuayb
  Zarar}.} \bibinfo{year}{2018}\natexlab{}.
\newblock \showarticletitle{Machine learning for placement-insensitive inertial
  motion capture}. In \bibinfo{booktitle}{\emph{2018 IEEE International
  Conference on Robotics and Automation (ICRA)}}. IEEE,
  \bibinfo{pages}{6716--6721}.
\newblock


\bibitem[Yirmibesoglu and Menguc(2016)]%
        {yirmibesoglu2016hybrid}
\bibfield{author}{\bibinfo{person}{Osman~Dogan Yirmibesoglu} {and}
  \bibinfo{person}{Yigit Menguc}.} \bibinfo{year}{2016}\natexlab{}.
\newblock \showarticletitle{Hybrid soft sensor with embedded IMUs to measure
  motion}. In \bibinfo{booktitle}{\emph{2016 IEEE International Conference on
  Automation Science and Engineering (CASE)}}. IEEE, \bibinfo{pages}{798--804}.
\newblock


\bibitem[Zhang et~al\mbox{.}(2015)]%
        {zhang2015whole}
\bibfield{author}{\bibinfo{person}{Yizhai Zhang}, \bibinfo{person}{Kuo Chen},
  \bibinfo{person}{Jingang Yi}, \bibinfo{person}{Tao Liu}, {and}
  \bibinfo{person}{Quan Pan}.} \bibinfo{year}{2015}\natexlab{}.
\newblock \showarticletitle{Whole-body pose estimation in human bicycle riding
  using a small set of wearable sensors}.
\newblock \bibinfo{journal}{\emph{IEEE/ASME Transactions on Mechatronics}}
  \bibinfo{volume}{21}, \bibinfo{number}{1} (\bibinfo{year}{2015}),
  \bibinfo{pages}{163--174}.
\newblock


\bibitem[Zhu et~al\mbox{.}(2021)]%
        {zhu2021robust}
\bibfield{author}{\bibinfo{person}{Zhongguan Zhu}, \bibinfo{person}{Shihui
  Guo}, \bibinfo{person}{Yipeng Qin}, \bibinfo{person}{Xiaowei Chen},
  \bibinfo{person}{Ronghui Wu}, \bibinfo{person}{Yating Shi},
  \bibinfo{person}{Xiangyang Liu}, {and} \bibinfo{person}{Minghong Liao}.}
  \bibinfo{year}{2021}\natexlab{}.
\newblock \showarticletitle{Robust elbow angle prediction with aging soft
  sensors via output-level domain adaptation}.
\newblock \bibinfo{journal}{\emph{IEEE Sensors Journal}} \bibinfo{volume}{21},
  \bibinfo{number}{20} (\bibinfo{year}{2021}), \bibinfo{pages}{22976--22984}.
\newblock


\end{thebibliography}

\appendix
\section{Implementation Details}

\cg{\begin{itemize}
    \item Our method: Except for some small-range-limited hyperparameters, we started with a rough search and then fine-tuned those hyperparameters as follows:
we varied the number of LSTM layers from
$[3, 4, 5, 6, 7]$, the number of units for the LSTM layer from
$[64,128, 256, 512]$, the batch size from $[128, 256,
512, 1024]$, $\eta$ of Equation~\ref{equ:totalloss} from
$[3000000, 4000000, 5000000, 6000000, 7000000]$, the m of Equation~\ref{equ:lambda1} from $4.69 \times 10e-4$
$[9.1\times10e^{7}, 1\times10e^{8}, 1.01\times10e^{8}, 1.02\times10e^{8}]$, the values of the $r$ of Equation~\ref{E1}, the learning rate from $[1e-1,1e-2,1e-3,
1e-4]$, and the ratio of predict error and measure error of the error-state Kalman filter from $[1.33,2,2.67,3.33,4]$. For the value of $r$ and m that have limited range, we varied $r$ from $[0.1, 0.15,
0.2, 0.25]$ and m from $[2,3]$ At last, we set the $\eta$ of Equation~\ref{equ:totalloss} as 5000000, the m of Equation~\ref{equ:lambda1} and $lambda$ of Equation~\ref{equ:lambda2} to 1010000000, the $r$ and $m$ of Equation~\ref{E1} to 0.25 and 2, respectively. Besides, we compared the result of transferring in the feature layer or output layer.
\item CL: This network consists of three sub-modules: (1) an embedding
layer consisting of four 1-dimensional convolution filters to learn embedding (the number of hidden units is set to 64); (2) an encoder
consisting of one long short-term memory (LSTM) layer (the number of LSTM units is set to 256); and
(3) an attention module consisting of a self-attention layer. The learning rate is 0.01. The hyperparameters and network structure were optimally chosen.
    \item LightGBM: The setting of parameters can refer to Table~\ref{tab:lgbparamset}. The hyperparameters and network structure were optimally chosen. 
    \item FCN: This model contains four hidden layers, using a varying number of $[128, 256, 512, 256]$ nodes in each layer. The learning rate is set to 0.01. The hyperparameters and network structure were optimally chosen.
    \item Random Forest: We set hyperparameters as follows:
    max depth to 3, n estimators to 150 and min samples split
    to 3. The hyperparameters and network structure were
    optimally chosen.

\end{itemize}

\begin{table}[h]
    \centering
    \caption{Patameters of LightGBM.}~\label{tab:lgbparamset}
    \begin{tabular}{|c|c|}
       \hline
	    Setting &Value  \\
 		\hline
 		num leaves  &100\\
 		\hline 
       max bin&76\\
       \hline 
       max depth& 19\\
       \hline 
       boosting type& gbdt\\
       \hline 
       reg alpha& 0.0001\\
       \hline 
       reg lambda& 0.0001\\
       \hline 
       min data in leaf& 5\\
       \hline 
       learning rate& 0.09\\
       \hline 
       feature fraction& 0.7 \\
       \hline 
       bagging fraction& 0.7  \\
       \hline
       num trees&60\\
       \hline
    \end{tabular}
      
\end{table}}

\end{document}